\documentclass[aps,onecolumn,groupedaddress,floatfix,nofootinbib]{revtex4-1}

\usepackage[colorlinks=true,citecolor=red,filecolor=green,linkcolor=blue,pdfnewwindow=true]{hyperref}
\usepackage{amsmath} 
\usepackage{amssymb}
\usepackage[version=4]{mhchem}
\usepackage{color}
\usepackage{graphicx}
\usepackage{makeidx}
\usepackage{bm}
\usepackage[title]{appendix} 

\usepackage{float}
\usepackage{caption}
\usepackage{hyperref}
\usepackage[normalem]{ulem}
\usepackage{natbib}
\usepackage[dvipsnames]{xcolor}
\usepackage[english]{babel}
\usepackage[autostyle]{csquotes}
\usepackage{algorithm} 
\usepackage{algpseudocode}
\usepackage{epigraph}
\usepackage{multirow}

\hypersetup{urlcolor=blue}

\makeatletter
 
\newcommand{\Rmnum}[1]{\expandafter\@slowromancap\romannumeral #1@}
\makeatother

\begin{document}
\title{Network based control strategies\\ for sustainable management of \textit{Lantana camara} }

\author{Shyam Kumar}
\thanks{These two authors contributed equally}
\author{Preet Mishra}
\thanks{These two authors contributed equally}
\author{R.~K.~Brojen Singh}
\email{Corresponding author: brojen@jnu.ac.in}
\affiliation{School of Computational $\&$ Integrative Sciences, Jawaharlal Nehru University, New Delhi-110067, India.}

\begin{abstract}
	
\noindent  
Control studies in ecological models of networks involving invasive plants can offer tangible management strategies. Framing control policies to tackle problems posed by invasive plants on indigenous plant diversity under the sustainable developmental goals framework is the main motivation of this paper.  Based on the reported experimental and theoretical observations, we propose minimal ecological models of direct interactions of \textit{Lantana camara} with other entitites e.g. other indigenous plants (denoted by Control-plant) and soil microbes. The species' abundance variables represented by the nodes in the model networks follow generalized Lotka-Volterra dynamics and input signals are incorporated at appropriate driver nodes to control the \textit{Lantana camara} outgrowth. Analysis with Lie algebra and controllability criteria showed that our model systems are controllable with minimum of two control input signals. Analytical results of our model equations indicated that successive interventions are required to increase the abundance of control-plant, the microbes and suppress \textit{Lantana camara}. We observed from the numerical analysis with LQR algorithm that the number of interventions to be taken to control the \textit{Lantana camara} outgrowth depends on the choice of various important parameters such as intervention time, effort and penalty etc. Further, the increase in nonlinear terms contributed from self-loops in the network model shows the persistence of \textit{L. camara} and how it drives the coexistence of the species. We propose that parameters we have identified in our study needed to be considered and analyzed extensively for designing sustainable policies to control invasive species in an ecosystem.\\

{\noindent}{{\it \textbf{Keywords:}} \textbf{Invasive species, \textit{Lantana camara}, Non-linear dynamics, Ecological Networks, Control theory, UN-SDG}}

\end{abstract}

\maketitle

\tableofcontents
\section{Introduction}
{\noindent}Sustainable development \cite{UNSDG} based on inclusive environmental, socio-economical policies which includes maintaining species diversity is identified as one of the major challenges we face today. The United Nations(UN), recognising the need to cope with changing conditions on the planet, has recommended a clear and important mandate of implementing Sustainable Development Goals (SDGs) \cite{United}. Hence, the conservation of local plant diversity particularly indigenous species of plants and trees has emerged as one of these goals.\cite{UNSDG}. Thus, the major aim can be defined as to maintain species diversity through proper control mechanism \cite{Harris, Alexander} by designing and implementing models based on species population obeying classical Malthusian growth \cite{Malthus}. Though the aim to conserve plant species is simplistic at first sight, this has a complex web of inter dependencies in the form of the ecological agents that we have to deal with. One such problem is controlling invasive species while designing plant diversity conservation policies. In biological control programs eco-friendly labelled solutions, etc. are not in general sustainable because many-a-time they lack thorough theoretical foundations. Any wrong policy decision can introduce extra loads on an already strained and invaded delicate eco-system. Thus, factoring in sustainability we have to design control programs based on strong fundamental theoretical principles. \\
\begin{figure*}[htbp]
    \centering
    \includegraphics[scale=0.5]{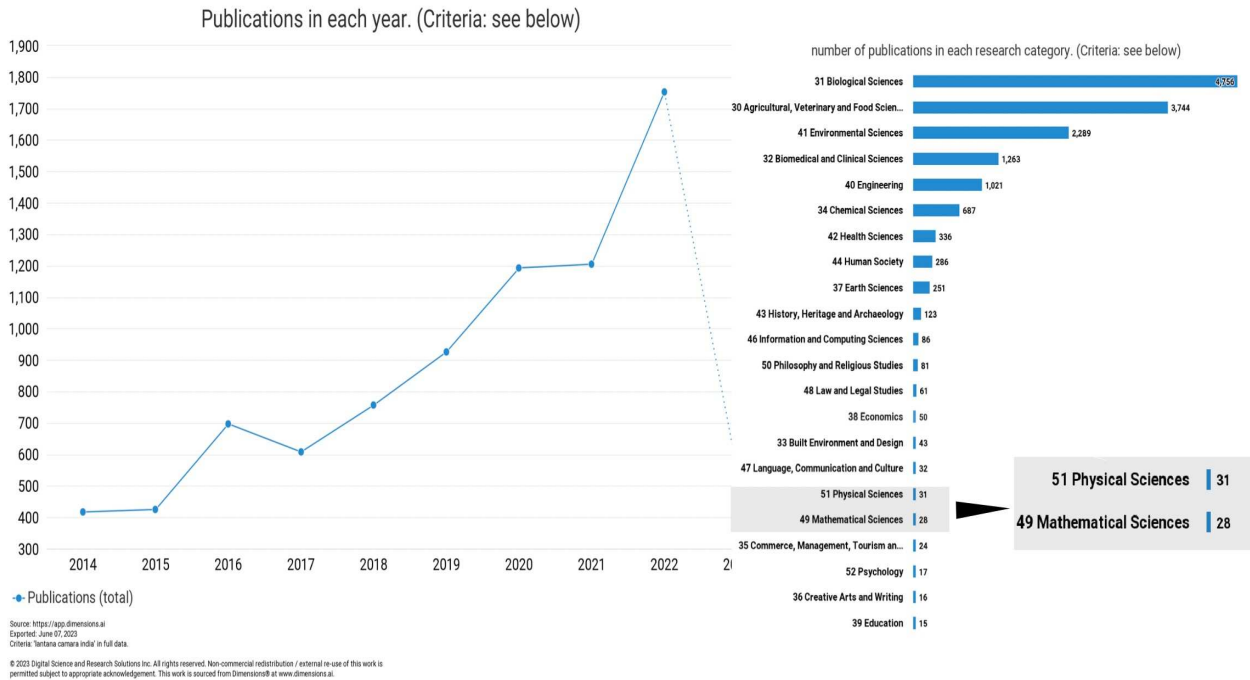}
    \caption{ Bibliometric meta data from Dimensions.ai. The highlighted box shows the number of publications in physical sciences and mathematics. This shows very little mathematical modelling work has been done in  management of invasive plants. 
    \label{fig1}}
\end{figure*}

\noindent\textit{Lantana camara} is one of the most invasive species currently spreading in India. The way \textit{Lantana camara} is moving across Indian landscapes has been repeatedly appeared in various research papers and gives a grim threatening of the future situation of the species diversity \cite{Sivakumar2018, Garkoti2021, Chauhan2022, Kohli2006, Negi2019, Vasudevan1991, Hiremath2018,Sharma2000, Joshi1991,Sundaram2012, Ramaswami2014, Love2009,Flory2009,Babu2009}. To provide a context of the matter studied and method adopted in this paper we have provided a  bibliometric analysis of the meta-data from the Dimensions.ai (accessed on June 07, 2023 ) using the search term ["\textit{Lantana camara }India"] as shown in the Fig. \ref{fig1}. This search term is by no means final and can be expanded to include various other terms. The goal of choosing the particular query term used in this paper was to find out an overview in a broad sweep the publications on this species which are directly related to the Indian diaspora. The results clearly show that the studies on \textit{Lantana camara} by researchers are increasing on an average with time in years (Fig. \ref{fig1} left panel) and the studies are done from wide spread disciplines listed in the Fig. \ref{fig1} right panel. Though we can see from the Fig. \ref{fig1} a steady increase in number of publications with year, the subject areas in which these publications are made shows a gap (very few papers from physical and mathematical sciences perspectives) present in the domain. The results indicate that little work has been done to mathematically model control policies that are implemented in this situation of \textit{Lantana camara}. Hence, we need to study the possible control strategies of this invasive species in a certain demographic region which can be extended to other species as well for possible sustainable development policies.
 \begin{figure}[H]
    \centering
    \includegraphics[scale=0.4]{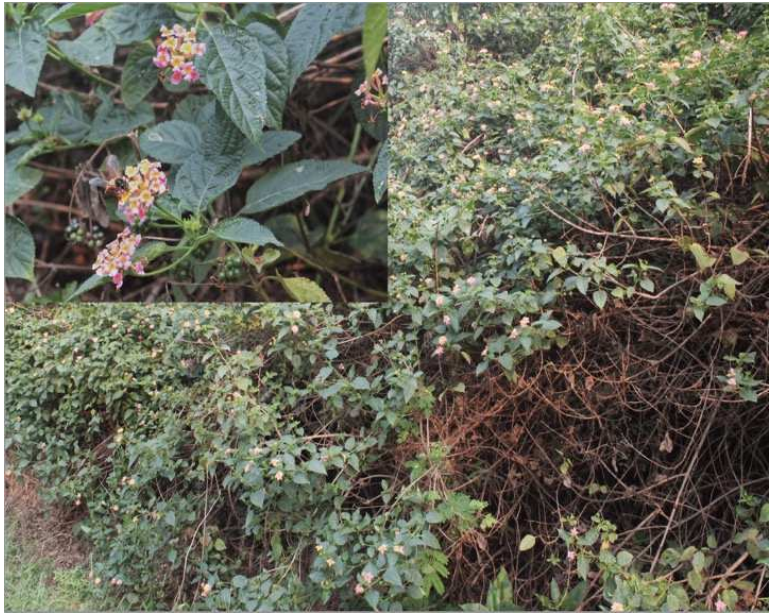}
    \caption{\textit{ Lantana camara }in an invaded landscape.}
\end{figure} 
\noindent   Various optimal policies has been proposed in the past (a sample set of which can be found in \cite{Vasudevan1991, Joshi1991, Ramaswami2014, Babu2009}).Majority of the policies implemented have emphasised on community participation, agrarian solutions, usage of bio-control solutions and usage of the invasive plant itself for commercial purposes to keep it under control. Though these plans are economic and time-optimal, they fail to capture the dynamic response of the other ecological agents present together with \textit{Lantana camara} in the field of operations during the plan implementation epoch. Indian landscapes have varied needs hence we have designed our strategy keeping in tune with UN-SDGs. Controlling ecological entities, specially invading species, in a sustainable way needs to be done in a systematic and strategic way as the population growths of these species depend on various ecological factors, climates, demographic parameters etc, and has to be correlated with any action plan to be implemented \cite{Weidlich}. The invasive plant, \textit{Lantana camara} has relatively very little positive impacts on the indigenous ecosystem.  It has certain anthropological usages such as, it can be used as garden ornament, firewoods, fertilizers, paper fibers, etc \cite{Negi} mosquito repellent \cite{Sharma}, has multi-ingredient medicinal valued as anti-fungal, anti-bacterial, anti-insecticidal \cite{Arunkumar}. Hence, making trade-offs based on cost-benefit scenarios are also needed to be factored into the control policy decision making programs. This gives us an important insight that we cannot completely remove \textit{Lantana camara} from the landscape. \\

{\noindent}Components of an ecosystem, such as, plants, animals, soil microbes etc. interact in a complex manner and can be represented through a network. In such networks, ecological agents are represented by the nodes (\textbf{V}) and interactions by edges (\textbf{E}) among them \cite{Cohen1968, Roberts1978}. Thus modelling of the time evolution of this ecosystem has multi-variable character and has hierarchical levels in system's organization \cite{Pickett}. The qualitative understanding of the non-linear dynamical system and the emergent after-effects of the non-linearity in the dynamics of these ecological agents are very important for designing preventive or curative actions in conservation and sustainable policies. Modelling such components with simplistic models by representing them with dynamical systems approach \cite{Mambuca} where the state of the ecological system at any instant of time 't' can be represented by species concentration or abundance vector $\textbf{x(t)}=[x_1,...,x_n]$, where, $x_i$ is the abundance of the ith species. The dynamical behaviour of the state vector $\textbf{x}(t)$ is then given in the most general case by,
\begin{eqnarray}
\label{intro}
\frac{d\textbf{X(t)}}{dt} = \textbf{F}({\bf X(t)},t)=\left[\begin{matrix}F_1\\F_2\\ \vdots \\F_n  \end{matrix}\right];~~F_i=F_i(x_1(t),x_2(t),...,x_n(t),t),~~\bf{X}=\left[\begin{matrix}x_1(t)\\x_2(t)\\ \vdots \\x_n(t)  \end{matrix}\right]
\end{eqnarray}
$F_i(x_1,..,x_n;t)$s can generally be inferred from experimental data and often found to be highly non-linear character \cite{May1975,Maier2013}. In the case of complex ecological networks, these functions $F_i;i=1,2,...,n$ can be represented with the network adjacency matrix $A=[A_{ij}]_{n\times n};i,j=1,2,...,n$ which characterize species interaction types and topologies \cite{Ulanowicz}. Once we construct the network dynamical system, we then ask how can we steer the system in finite amount of time to achieve a certain desired state by suitable inputs to particular nodes (ecological agents) of the network. For the analysis and solution to this situation, network theory based controllability \cite{Angulo2019,Liu2011,Lin1974,Liu2016} predictions have been systematically considered in this work.\\

{\noindent}In the context of designing sustainable control policies we must plan the steps in such a way that it will not only manage the control of \textit{Lantana camara} invasion but also replenish the soil (which has been depleted due to invasion) , generate some useful products or create positive impacts due to the presence of control plants (C-plants) for the communities which live in that niche. Adhering to the UN SDG\cite{UNSDG} programs, we have taken together non-linear dynamics of ecological systems, network controllability theory and developed a framework that can provide robust sustainable solutions for management of invasive species when implemented in the real field operations. To formulate and show the working principle of the possible derived policies through the  developed framework, we have presented two variants (Model 1 and Model 2) in the case study of managing \textit{Lantana camara} and their analysis. We also tried to show that the theoretical basis of the work is to help in developing and implementing of such derived policies in both quantitative and qualitative sense.  The only constraints of such policies being labour consistency, how much we really care for the environment and how efficiently we can manage our future footprints. \\
\begin{figure*}[htbp]
    \centering
    \includegraphics[scale=0.5]{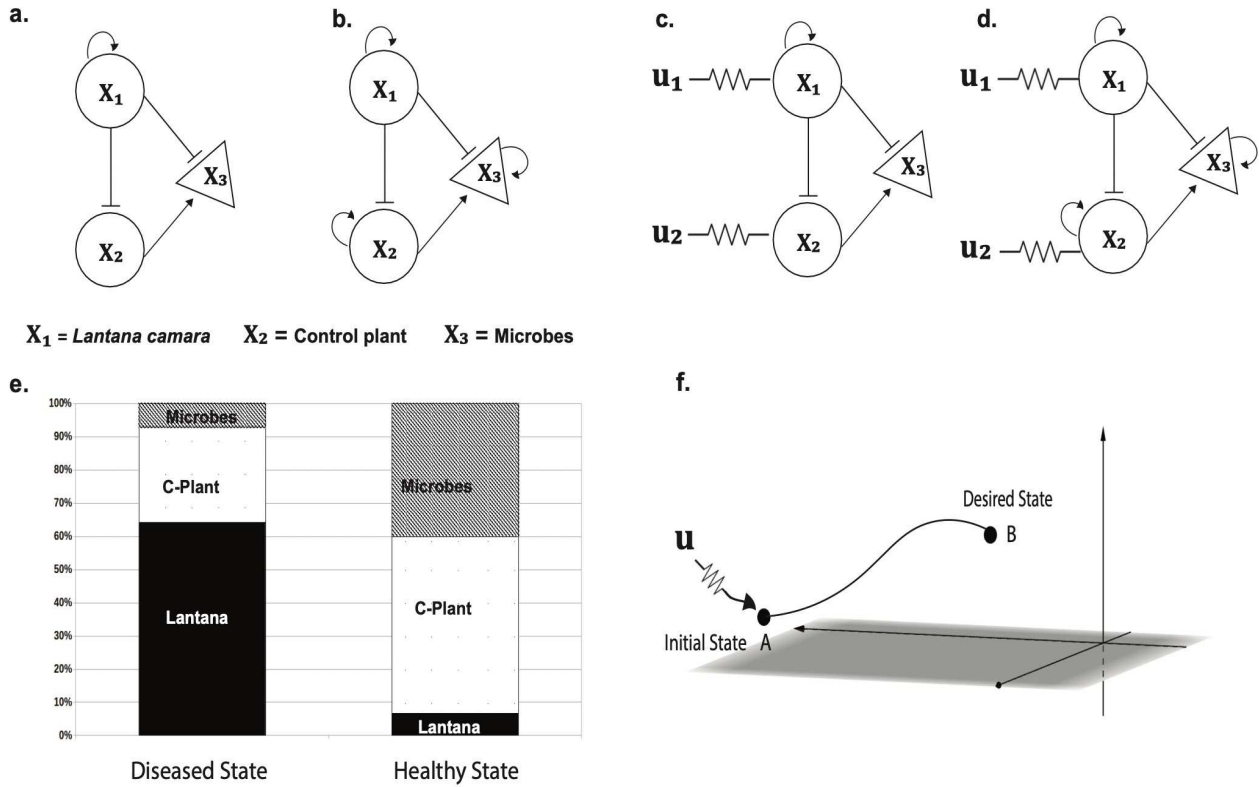}
    \caption{ (a) and (b) shows the network model of \textit{Lantana camara}($X_1$), Control plant($X_2$) and microbes($X_3$). (a) correponds to Model-1 where the self loop is present only in($X_1$). (b) corresponds to Model-2 where all nodes have self loops. (c) and (d) shows the controlled network with inputs $u_1$ and $u_2$ in appropriate driver nodes. (e) We have defined two states of the ecological system. Diseased state which shows an abnormal abundance of Lantana camara, which results in decreased presence of Control plant and Microbes. Healthy state where \textit{Lantana camara} is kept at a minimum  (f) Steering the system from an initial diseased state to a desired healthy state on a manifold using control input signals \textbf{u}.  }
    \label{fig3}
\end{figure*}
\section{Models and Methods}
\subsection{Minimal models of \textit{Lantana camara} with control strategies}
{\noindent}Ecosystem can be represented by a complex network of diverse interacting species and is quite vulnerable to intrinsic and extrinsic perturbations causing fragile behaviours in the ecological dynamics \cite{Montoya} which could be sometimes restorable \cite{Jones} whereas some are quite difficult to repair \cite{Singh}. The behaviour of such ecological systems can be well studied within the framework of population dynamics which study the how the population state of species of the system changes over time due to interactions within individuals of the same species and with individuals of other species present in the ecosystem. Further, by using population dynamics models, we can make predictions about the future state of the species. If we know the dynamical behaviour of the system, any arbitrary state can be reached from any specified state through external actions i.e. by application of suitable control strategy
\cite{Angulo2019,Liu2016,Liu2011,Lin1974,Wei1963,Conte1999,Brockett2014,Brockett2015,Brockett1973,Nijmeijer1990,Olver2000,Costanza1997,Costanza1990}.\\

{\noindent} The invasive species \textit{Lantana camara} generally spreads out rapidly and has caused harm to other indigenous plant species reducing the species diversity \cite{Hejda}, causing species extinction \cite{Keyes}, and is one of main drivers of imbalance between local-global diversity \cite{Pysek}. This has resulted in a state of the ecosystem which we have termed as diseased state. Thus the aim of our work is, by using suitable external inputs, to guide the diseased ecosystem to a desired healthy state. Furthermore, we have two-fold objectives, first we want to manage the abundance of \textit{Lantana camara} population and second, we want to preserve the healthy state (soil health, species diversity etc.). Since inter-species competition in plants can induce changes in the abundances of the plants inhibiting within threshold physical distances of each other in the ecosystem \cite{Joshi1991,Sakai1954}, we took this plant-plant interaction in our model. Hence, we introduce \textit{Control-plant} (C-plant) interacting with \textit{Lantana camara} in our minimal model to incorporate the mentioned factor for controlling the abundance of \textit{Lantana camara}.\\

{\noindent}For the second objective, we need to incorporate in the model how an ensemble of plants effect the biotic components of the soil i.e. microbes and vice versa \cite{Doran}. In general, any realistic systems representation of such ecological dynamics involve a dense complex interaction of multi-scaled components, hierarchies and dynamics \cite{Poole}, which we simplify to a basic model to capture relevant features of the ecological dynamics. Further, there is clear evidence of dependence of plant fitness, productivity and plant disease reduction on the interacting microbial communities in the soil which manage its' vital conditions \cite{Waldrop,Pascale} and plant-microbial interaction, where, the microbe-microbe interaction structures/restructures the topology of the microbial communities \cite{Bakker,Hassani}. Hence, we consider \textit{Lantana camara}-microbe interaction and control plant- microbes interaction in minimal models which could play an important role to control soil quality. This is what we define as sustainable controls.  Our models are not specific only for \textit{Lantana camara} but could be applicable to any invasive plants modelling in any ecosystem.\\

{\noindent}We have constructed two variants of the network between \textit{Lantana camara}, the C-plant and the soil microbes as can be seen in Fig\ref{fig3} a, b. We have not assigned any particular plant as C-plant in the outset as its selection can be balanced between local as well as global needs depending upon variety of factors. We have constructed the interactions in the networks among the three species by the following rules:
1. \textit{Lantana camara} inhibits C-plant nearby \cite{Kato} and soil microbes \cite{SinghHP}, 2. C-plant promotes the soil microbes in general \cite{Hamilton,Wardle}. Thus, the requirements while selecting the C-plant is that it follows the above rules. This is based on the observation that \textit{Lantana camara} promotes the growth of deleterious microbes that decrease the population of beneficial microbes in the soil \cite{Garkoti2021, Fan2010} and implicitly hinder the growth of other plants. Here we make a suggestion about keeping the choice of the C-plant as Leguminous plants. This has its own benefits as these plants are known to improve soil quality by increasing and promoting the growth of good microbes.\\

\subsection{Dynamics of ecological networks}
{\noindent}Network of $n$ ecological agents can be defined by a graph $G(N,E)$, where, $N=\{x_i;i=1,2,...,n\}$ as the set of nodes, and all the possible interaction between any pair of nodes in the network is defined by the set of edges in the graph, $E=\{e_{ij};\forall i,j\in N\}$. The topological features of the network is characterized by the adjacency matrix $A=[A_{ij}]$ of the network, where, the matrix elements $A_{ij}$ can be defined by,
\begin{eqnarray}
A_{ij}=\begin{cases}
1 ,&~~\text{if ith and jth nodes have an edge  } \\
0,&~~\text{otherwise}
\end{cases}
\end{eqnarray}
We want to study the dynamics of the network of path length one, we need to know the degree of each node $i$ $x_i$ which is $k_i=\displaystyle\sum_{<i,j>}^n A_{ij}$, where, $<...>$ indicates nearest neighbour to ith node, where, $A_{ij}\ne 0$. For $A_{ij}\ne 0,\forall i\ne j$, such that, there exist $k_i$ such $(x_j\rightarrow x_i)\in E$ edges to the ith node.

In this case, the neighbouring nodes will serve as coupling nodes to the ith node, whereas, for $A_{ii}\ne 0$ will serve as intrinsic dynamics of the ith node.

Hence, there will be $k_i-1$ coupling terms given by coupled reactions, $lx_j+mx_i\displaystyle\stackrel{A_{ij}}{\longrightarrow}m^\prime x_i$ and one intrinsic reaction, $mx_i\displaystyle\stackrel{A_{ii}}{\longrightarrow}m^\prime x_i$, where, the possible conditions $m>m^\prime$ and $m<m^\prime$ indicate the decay or direct inhibition and creation or direct promotion in the growth of the species variable $x_i$ respectively. Then, the dynamics of ith variable at ith node can be obtained by applying simple mass action kinetics of the reactions involved with $x_i$ in the network, where, each contributing term in coupling becomes $A_{ij}H_{ij}(x_i,x_j);j=1,2,...,n$, such that,  the coupling function, $H_{ij}(x_i,x_j)=x_i^mx_j^l$ (generally $m\ge 1, l\ge 1$), whereas, the intrinsic contribution due to self-loop in the node is given by $A_{ii}H_{ii}(x_i)=A_{ii}x_i^m$ (generally $m\ge 2$). Linear network dynamics of path length one is given by the condition $l=1,m=0,m^\prime=1$ for coupling functions giving $H_{ij}(x_i,x_j)=x_j$ and $m=1,m^\prime=2$ for intrinsic contribution in the dynamics providing $H_{ii}(x_i)=x_i$. Further for the ith node there could be other terms apart from coupling and self-loop, which is spontaneous decay (may be natural death) $x_i\displaystyle\stackrel{d_{i}}{\longrightarrow}\phi$ or/and creation (formation or reproduction) of the variable $x_i\displaystyle\stackrel{r_{i}}{\longrightarrow}2x_i$, where, $d_{i}$ and $r_{i}$ are intrinsic death and growth rate of the species variable $x_i$ at ith node of the network. Hence, the dynamics of the network of path length one is given by \cite{Newman},
\begin{eqnarray}
\frac{dx_i}{dt}&=&\sum_{i,j=1}^{n}A_{ij}H_{ij}(x_i,x_j)-d_{i}x_{i}+r_{i}x_i\noindent\\
&=&\left[A_{ii}\frac{H_{ii}(x_i)}{x_i}-d_{i}+r_i\right]x_i+\sum_{<i,j>}^{n}A_{ij}H_{ij}(x_i,x_j)
\end{eqnarray} 
We can write the dynamics of the network as given by,
\begin{eqnarray}
\label{network}
\frac{d\textbf{X}}{dt}&=&D[\textbf{X}]\left[\textbf{B}\textbf{X}-\textbf{d}+\textbf{r}\right]\\
&&\textbf{B}=\left[\begin{matrix}B_{11}&B_{12}&\dots&B_{1n}\\
B_{21}&B_{22}&\dots&B_{2n}\\
\vdots&\vdots&\dots&\vdots\\
B_{n1}&B_{n2}&\dots&B_{nn}
\end{matrix}\right],~~
D[\textbf{X}]=\left[\begin{matrix}x_{1}&0&\dots&0\\
0&x_{2}&\dots&0\\
\vdots&\vdots&\dots&\vdots\\
0&0&\dots&x_{n}
\end{matrix}\right],~~
\textbf{X}=\left[\begin{matrix}x_1\\x_2\\ \vdots\\x_n\end{matrix}\right],~~
\textbf{d}=\left[\begin{matrix}d_1\\d_2\\ \vdots\\d_n\end{matrix}\right],~~
\textbf{r}=\left[\begin{matrix}r_1\\r_2\\ \vdots\\r_n\end{matrix}\right]\nonumber
\end{eqnarray}
where, $\textbf{X}(t)\in \mathbb{R}^n$ . The $(ij)th$ matrix element of the matrix $B$ is given by, $\displaystyle B_{ij}=\frac{A_{ij}H_{ij}(x_i,x_j)}{x_ix_j}$ and diagonal elements $(ii)th$ term is given by, $\displaystyle B_{ii}=\frac{A_{ii}H_{ii}(x_i)}{x_i^2}$. We can classify two major types of models as follows:
\begin{itemize}
\item\textbf{Linear dynamics}: In this case we have, for ith node, coupling term coming from reaction of type $x_j\displaystyle\stackrel{A_{ij}}{\longrightarrow}x_i$ becomes $A_{ij}x_j$ (following mass action kinetics) such that $H_{ij}=x_j$. In this situation, $\displaystyle B_{ij}=\frac{A_{ij}H_{ij}}{x_ix_j}=\frac{A_{ij}}{x_i}$, so that, $D[\textbf{X}]\textbf{B}=\textbf{A}$. Then, the dynamics of the network given by equation \eqref{network} becomes,
\begin{eqnarray}
\label{linear}
\frac{d\textbf{X}}{dt}&=&\textbf{A}\textbf{X}+D[\textbf{X}]\left[-\textbf{d}+\textbf{r}\right]\\
&&D[\textbf{X}]=\left[\begin{matrix}x_{1}&0&\dots&0\\
0&x_{2}&\dots&0\\
\vdots&\vdots&\dots&\vdots\\
0&0&\dots&x_{n}
\end{matrix}\right],~~
\textbf{X}=\left[\begin{matrix}x_1\\x_2\\ \vdots\\x_n\end{matrix}\right],~~
\textbf{d}=\left[\begin{matrix}d_1\\d_2\\ \vdots\\d_n\end{matrix}\right],~~
\textbf{r}=\left[\begin{matrix}r_1\\r_2\\ \vdots\\r_n\end{matrix}\right]\nonumber
\end{eqnarray}

\item\textbf{Generalized Lotka-Volterra (GLV) dynamics}: The simplest form of the the coupling function $H_{ij}$ of ith species variable represented by ith node of the network is given by $H_{ij}=x_ix_j$ contributed from the reaction $\displaystyle x_i+x_j\stackrel{A_{ij}}{\longrightarrow}2x_i$ (mass action kinetics). In this case, $\textbf{d}=\textbf{0}$. The matrix elements of \textbf{B} becomes, $\displaystyle B_{ij}=\frac{A_{ij}H_{ij}}{x_ix_j}=A_{ij}$ such that $\textbf{B}=\textbf{A}$. Now, the from equation \eqref{network}, the GLV network dynamics becomes,
\begin{eqnarray}
\label{glv}
\frac{d\textbf{X}}{dt}&=&D[\textbf{X}]\left[\textbf{A}\textbf{X}+\textbf{r}\right]\\
&&D[\textbf{X}]=\left[\begin{matrix}x_{1}&0&\dots&0\\
0&x_{2}&\dots&0\\
\vdots&\vdots&\dots&\vdots\\
0&0&\dots&x_{n}
\end{matrix}\right],~~
\textbf{X}=\left[\begin{matrix}x_1\\x_2\\ \vdots\\x_n\end{matrix}\right],~~
\textbf{d}=\left[\begin{matrix}0\\0\\ \vdots\\0\end{matrix}\right],~~
\textbf{r}=\left[\begin{matrix}r_1\\r_2\\ \vdots\\r_n\end{matrix}\right]\nonumber
\end{eqnarray}
The functional form of $H_{ij}$ is very complicated which could be various forms of complex response functions imparted from the coupled neighbouring nodes of path length one \cite{Angulo2019}. The general functional form which could be derived from various forms of the response functions given in \cite{Jost} is given by,
\begin{eqnarray}
H_{ij}(x_i,x_j)=\frac{a_{ij}x_ix_j^{\alpha}}{(1+a_{ij}h_{ij}x_j^{\beta})(1+c_{ij}x_i^{\gamma})}
\end{eqnarray}
\begin{itemize}
\item If $\alpha=1, \beta=1, \gamma=0$, $H_{ij}$ is Holling type-II.
\item If $\alpha=2, \beta=2, \gamma=0$, $H_{ij}$ is Holling type-III.
\item If $\alpha=1, \beta=1, \gamma=1$, $H_{ij}$ is Crowley-Martin type.
\item If $\alpha=1, \beta=1, \gamma=1$ and neglecting $O(x^{\beta+\gamma})$, $H_{ij}$ is De-Angelis-Beddington type.
\item If $\alpha=1, \beta=1, c_{ij}=1, x_i<<x_j, H_{ij}$ is Hassel-Varley type-I.
\item If $\alpha=1, \beta=1, c_{ij}=1, O(x^{\beta+\gamma}) <<1<<x_i^{\gamma}+a_{ij}h_{ij}x_j$, $H_{ij}$ is Hassel-Varley type-II.
\item $If \alpha=1, \beta=1, \gamma=1, c_{ij}=1$ and $O(x^{\beta+\gamma})<1<<x_i+a_{ij}h_{ij}x_j$, $H_{ij}$ is Ratio dependent type II.
\item If $\alpha=2, \beta=2, \gamma=2, c_{ij}=1$ and $O(x^{\beta+\alpha})<1<<x_i^2+a_{ij}h_{ij}x_j^2$, $H_{ij}$ is Ratio dependent type III.
\end{itemize}
where, $a_{ij}$, $h_{ij}$ and $c_{ij}$ are the attack rate, handling time and conversion efficiency of the jth species to ith species sitting at ith and jth nodes of the network respectively. $\alpha$, $\beta$ and $\gamma$ are the constant exponents of the species variables. The intrinsic functional form of the GLV $H_{ii}$ is given by \cite{Angulo2019,Jost},
\begin{eqnarray}
H_{ii}(x_i)=x_i\left[r_i+\left(1-\frac{x_i}{w_i}\right)\left(\frac{x_i}{K_i}-1\right)\right]
\end{eqnarray}
where, $r_i$, $w_i$ and $K_i$ are growth rate, allele effect and carrying capacity of the ith species variable respectively.
\end{itemize}

{\noindent}Now, let us consider our minimal model network of invasive plant \textit{Lantana camara}, C-plant and soil microbes represented by variables $x_1$, $x_2$ and $x_3$ respectively (Fig.\ref{fig3}, a, b). We used generalized Lotka-Volterra (GLV)  \cite{Goel}model equations to design the dynamics of each variable in the two models given in the Fig. 3 a, b. The time evolution of interaction among these three species i.e. $n=3$ which are represented by state vector $\textbf{X}(t) =[x_1~x_2~ x_3]^T \in \mathbb{R}^3$, where, $T$ is the transpose of the given vector, is governed by the coupled differential equations \eqref{network}, \eqref{glv} 
\begin{eqnarray}
\label{glv}
\frac{d\textbf{X}(t)}{dt}=D[\textbf{X}]\left[\textbf{A}\textbf{X}+\textbf{r}\right],~~~\textbf{A}=\left[\begin{matrix}A_{11}&A_{12}&A_{13}\\
A_{21}&A_{22}&A_{23}\\
A_{31}&A_{32}&A_{33}
\end{matrix}\right],~~
D[\textbf{X}]=\left[\begin{matrix}x_{1}&0&0\\
0&x_{2}&0\\
0&0&x_{3}
\end{matrix}\right],~~
\textbf{X}=\left[\begin{matrix}x_1\\x_2\\x_3\end{matrix}\right]~~
\textbf{r}=\left[\begin{matrix}r_1\\r_2\\r_3\end{matrix}\right]
\end{eqnarray}
All $A_{ij} $'s are positive real numbers. We use this GLV model equations in the two variants of the minimal network models in Fig.\ref{fig3} a, b.\\

\begin{figure*}[htbp]
    \centering
    \includegraphics[scale=0.35]{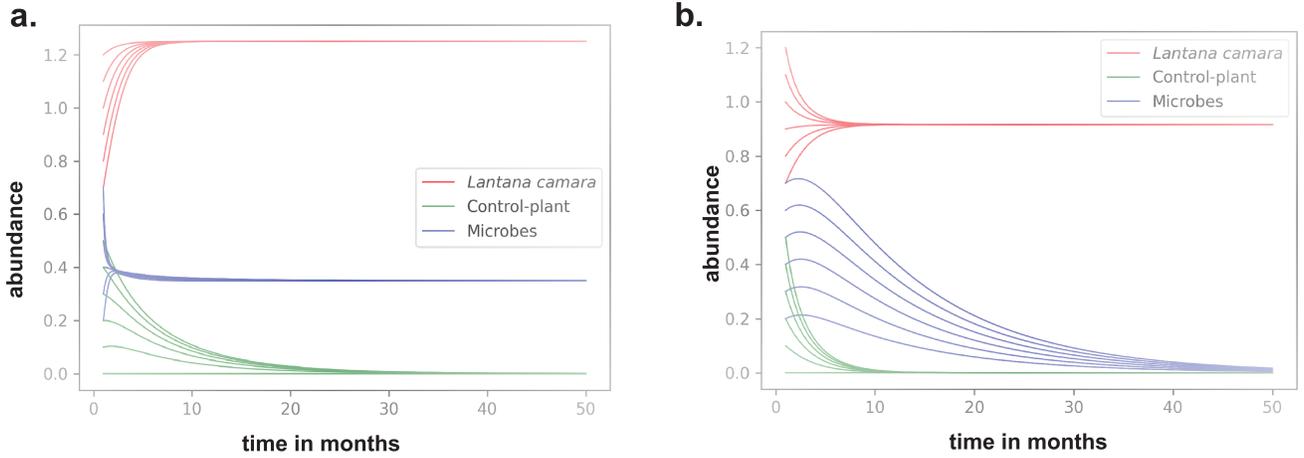}
    \caption{ RK-4 simulations of dynamics for two models in absence of control inputs (a) Model 1 refer equations:\ref{model1} and (b) Model 2 refer equations \ref{model2} }
    \label{fig4}
\end{figure*}

\subsubsection{Model 1}

{\noindent}This model considers intra-species interaction represented by self-loops in the network. We have the following assumptions for mathematical convenience and construction of the simplest possible non-linear model. The impact of intra-competition interactions are relatively higher amount and visible in \textit{Lantana camara} hence we have kept the interaction term $A_{11} \neq 0$ and of inhibitory type.  The self-loops are not considered in the two variables for the C-plants and microbes ($A_{22}, A_{33} = 0$) as their impact is magnitude lesser than as in \textit{Lantana camara}. For this model, the dynamical system given by \eqref{glv} reduces to the following equations,
\begin{eqnarray}
\label{model1}
\frac{d\textbf{X}}{dt}=D[\textbf{X}]\left[\textbf{A}\textbf{X}+\textbf{r}\right],~~\textbf{A}=\left[\begin{matrix}-A_{11}&0&0\\
-A_{21}&0&0\\
-A_{31}&A_{32}&0
\end{matrix}\right],~~
D[\textbf{X}]=\left[\begin{matrix}x_{1}&0&0\\
0&x_{2}&0\\
0&0&x_{3}
\end{matrix}\right],~~
\textbf{X}=\left[\begin{matrix}x_1\\x_2\\x_3\end{matrix}\right]~~
\textbf{r}=\left[\begin{matrix}r_1\\r_2\\r_3\end{matrix}\right]
\end{eqnarray}
where, $A_{21},A_{31}$, and $A_{32}$ are network coupling parameters and intrinsic parameter contributed from the self-loop is $A_{11}$.   The signs indicate whether the coupling is inhibitory or activator types. Then, from the network model-1 in Fig. \ref{fig3}a represented by the equations \eqref{model1}, the dynamics of the variables in the GLV model can be written as,
\begin{eqnarray}
\frac{dx_1}{dt}&=&-A_{11}x_1^2+r_1x_1 \label{m11}\\
\frac{dx_2}{dt}&=&-A_{21}x_1x_2+r_2x_2\\
\frac{dx_3}{dt}&=&-A_{31}x_1x_3+A_{32}x_2x_3+r_3x_3
\end{eqnarray}
In this model, the values used for $A_{ij}$ are arbitrary and indicative of assumptions about magnitudes of relative strengths of interactions of the agents when made on similar time-scales. This assumption also provides us the model equations with a lesser degree of non-linearity. 

\subsubsection{Model 2}
{\noindent}In this variant of the model, the degree of non-linearity is increased by allowing self-loops in all the nodes in the network as shown in Fig. \ref{fig3}b. The network coupled equation representing the model 2 can be obtained by the corresponding equation given by equation \eqref{network} as follows,
 \begin{eqnarray}
\label{model2}
\frac{d\textbf{X}}{dt}=D[\textbf{X}]\left[\textbf{A}\textbf{X}+\textbf{r}\right],~~\textbf{A}=\left[\begin{matrix}-A_{11}&0&0\\
-A_{21}&-A_{22}&0\\
-A_{31}&A_{32}&-A_{33}
\end{matrix}\right],~~
D[\textbf{X}]=\left[\begin{matrix}x_{1}&0&0\\
0&x_{2}&0\\
0&0&x_{3}
\end{matrix}\right],~~
\textbf{X}=\left[\begin{matrix}x_1\\x_2\\x_3\end{matrix}\right]~~
\textbf{r}=\left[\begin{matrix}r_1\\r_2\\r_3\end{matrix}\right]
\end{eqnarray}
where, $A_{21}, A_{31}$ and $A_{32}$ are network coupling parameters, and $A_{11},~A_{22}$ and$A_{33}$ are intrinsic parameters contributed from the three self-loops in the network model. \textit{Lantana camara} starts interacting among its' population only as indicated by its own production and self-loop in the network \eqref{m11}. Then it's interaction with the other species in the network, namely, C-plant and microbe community, start working through these species indicated by coupling terms and self-loops in the network. This condition definitely increase the degree of nonlinearity in the dynamics. The equivalent GLV dynamical equations of the network are given by,
\begin{eqnarray}
\frac{dx_1}{dt}&=&-A_{11}x_1^2+r_1x_1\\ 
\frac{dx_2}{dt}&=&-A_{21}x_1x_2-A_{22}x_2^2+r_2x_2\\
\frac{dx_3}{dt}&=&-A_{31}x_1x_3+A_{32}x_2x_3-A_{33}x_3^2+r_3x_3
\end{eqnarray}
The role of nonlinearity in dynamical systems is quite important which can induce different forms of oscillations, chaos \cite{Hassell}, patterns of fitness of the involved species in the ecosystems affecting evolutionary characteristics \cite{Metz}, fractal and bifurcation scenarios in the species distributions \cite{Strogatz}, multistability and spontaneous symmetry breaking \cite{Miri} and so on. We would like to study the impact of nonlinearity in the dynamics of the invasive plant \textit{Lantana camara} and its interacting partners in the ecosystem. Since the number of nonlinear terms is allowed to increase through self-interaction of the involved species in the network via self-loops, the changes in the dynamics and patterns is induced by the species themselves spontaneously which will be important to study sustainability of the species in the network.\\
\subsection{Driving ecological states of choice to sustain diversity with \textit{Lantana camara}}

{\noindent} In the case of dynamics with invasive species, such as, \textit{Lantana camara}, the ecological state generally stay around \textit{disease states} such as described below and shown by numerical solutions (Fig \ref{fig4}) due to over-dominance by the invasive species. Such states, which could be one of the equilibrium states of the ecosystem, are not desirable states as the process disturbs species diversity. In order to capture such states with the first model, we took the following parameter values in the model equation \eqref{model1} as given below.
\begin{eqnarray}
A=\left[\begin{matrix}
-0.6 & 0 & 0 \\
-0.5 & 0 & 0 \\
-0.8 & 0.7 & 0
\end{matrix}\right] 
\end{eqnarray}
The values used for the matrix elements of $A$ are arbitrary and indicative of assumptions about magnitudes of relative strengths of plant-plant interactions and microbe-microbe interactions in communities if made on the similar time-scales of interaction with other interaction partners in the model. Since \textit{Lantana camara} has  a very high, abnormal abundance, we have taken the diseased state of the landscape as the initial ecological state of the system. The possible diseased state is given by $X_0$ by,
\begin{eqnarray}
\label{initial}
X_0=\left[\begin{matrix}
0.9 \\
0.1 \\
0.4
\end{matrix}\right] 
\end{eqnarray}
We assumed that this diseased state is a point of equilibrium of the given dynamics, hence, the growth rate vector can be calculated from equation \eqref{model1} by putting $\frac{d\textbf{X}}{dt}=0$
\begin{eqnarray}
 r=-A x_0=\left[\begin{matrix}
0.54 \\
0.45 \\
0.65
\end{matrix}\right]
\end{eqnarray}
This yields the GLV (coupled ODEs) system as given by,
\begin{subequations}
\begin{align}
\dot{x}_{1}&=-0.6 x_{1}^{2}+0.54 x_{1}  \label{m1_1}\\
\dot{x}_{2}&=-0.5 x_{1} x_{2}+0.45 x_{2} \label{m1_2}\\
\dot{x}_{3}&=-0.8 x_{1} x_{3}+0.7 x_{2} x_{3}+0.65 x_{3}\label{m1_3}
\end{align}
\end{subequations}
Similarly, for the second model (model 2), we took,
\begin{eqnarray}
A=\left[\begin{matrix}
-0.6 & 0 & 0 \\
-0.5 &-0.2 & 0 \\
-0.8 & 0.7 & -8
\end{matrix}\right] 
\end{eqnarray}
This yields the GLV equations for the model 2,
\begin{subequations}
\begin{align}
\dot{x}_{1}&=-0.6 x_{1}^{2}+0.54 x_{1} \label{m2_1}\\
\dot{x}_{2}&=-0.5 x_{1} x_{2}-0.2{x^2}_2+0.45 x_{2} \label{m2_2}\\
\dot{x}_{3}&=-0.8 x_{1} x_{3}+0.7 x_{2} x_{3}-8{x_3}^2+0.65 x_{3}\label{m2_3}
\end{align}
\end{subequations}
These two sets of equations \eqref{m1_1}, \eqref{m1_2} and \eqref{m1_3} and \eqref{m2_1}, \eqref{m2_2} and \eqref{m2_3} are the GLV equations of the two models. The simulations of the model equations are done using a standard Runge-Kutta (RK)-4 based numerical scheme for various initial conditions $X_0$ where one of them is given in \eqref{initial}. We can see that for a set of range of initial values, \textit{Lantana camara} continues to be the dominant species in terms of abundance (Fig.\ref{fig4} a and b). It also shows that the abundance of the beneficial microbes (blue curves) goes to negligible within certain long interval of time (Fig.\ref{fig4} b). We also reiterate that this dismal state of the system is an equilibrium state with respect to the parameter values chosen as the states change become constant i.e. independent of time after some interval of time and become independent of initial conditions also. It may so happen that if these values coincide with a real eco-system this equilibrium state will persist for long duration of time. We have confined ourselves within the determinstic modelling of dynamics of \textit{Lantana camara} in invaded landscapes as target of our strategies.\\

{\noindent}One important way to control the abnormal growth of the invasive plant \textit{Lantana camara} is incorporating control signals $u$ to the network \cite{Angulo2019,Brockett1973}. If we incorporate $m~(\le n)$ control signals $u_j:j=1,2,...,m$ to our model network \eqref{glv}, we have, the following control equation,
\begin{eqnarray}
\label{control}
\frac{d\textbf{X}(t)}{dt}&=&D[\textbf{X}]\left[\textbf{A}\textbf{X}+\textbf{r}\right]+\sum_{j=1}^{m}g_j(x_j)u_j(t)\nonumber\\
&=&D[\textbf{X}]\left[\textbf{A}\textbf{X}+\textbf{r}\right]+\textbf{GU}(t)\\
&&\textbf{G}=\left[g_{ij}\right]_{n\times m},~~
\textbf{U}=\left[\begin{matrix}u_1\\u_2\\u_3\end{matrix}\right]\nonumber
\end{eqnarray}
where $\textbf{G(X)} \in \mathbb{R} ^{n\times m}$. In our model (Fig. \ref{fig3}b), we used two control signals $u_1$ and $u_2$ to the abundances variables of \textit{Lantana camara}($x_1$)  and C-plant ($x_2$) respectively. The choice bases from the fact that $x_1$ and $x_2$ are the driver nodes. This choice comes from using the maximum matching algorithm through which we can identify suitable driver nodes at which control inputs can be applied.  Then, from the equation \eqref{control}, we have,
\begin{eqnarray}
\label{data}
\textbf{G}=\left[\begin{matrix}
1&0\\
0&1\\
0&0
\end{matrix}
\right];~~
\textbf{g}_1=\left[\begin{matrix}1\\0\\0\end{matrix}\right],~~
\textbf{g}_2=\left[\begin{matrix}0\\1\\0\end{matrix}\right],~~
\textbf{U}=\left[\begin{matrix}u_1\\u_2\end{matrix}\right]
\end{eqnarray}
where it was assumed that the susceptibilities ($g_{ij}$) of the species to input  are constant and equal to one.

In terms of impulsive control equation which we have used \ref{control} can be written as :
\begin{equation}
\dfrac{d\textbf{X}\left( t\right) }{dt}= D[\textbf{X}]\left[\textbf{A}\textbf{X}+\textbf{r}\right],\
\Delta \textbf{X(t)} = \textbf{GU}(t)
\end{equation}

\subsection{Finding input nodes for controlling network}
{\noindent}Invasive plant/s generally disturb ecosystem dynamics driving it to \textit{disease state} and force to stay around it \cite{Montoya,Hejda}. To restore or repair the features of the ecosystems it is essential to incorporate control inputs to the minimum set of driving nodes of the network of the ecological system to bring back the \textit{disease state} to the desired \textit{healthy state} (Fig. \ref{fig3}f). For achieving the above goal we first need to choose the suitable nodes for applying control inputs ($u$). This has been done by finding minimum set of driver nodes in the ecological network \cite{Angulo2019,Conte1999,Liu2011}. The set of driver nodes of any graph $G(V, E)$ can be obtained by making of its bipartite graph and using maximum matching algorithm \cite{Hopcroft,Liu2011}. This can be done by matching $m$ input nodes in $G$ as a set of pairwise non-adjacent edges, none of which are loops; that is, no two edges share common vertices \cite{Hopcroft}. A matching $m$ of a graph $G$ is maximal if every edge in $G$ has a non-empty intersection with at least one edge in $m$ \cite{Liu2011}. Using this maximum matching algorithm we find suitable driver nodes at which control inputs can be applied for the two models respectively as shown in Fig. \ref{fig3} (c, d).\\

\subsection{Theory of nonlinear control method}

{\noindent}Modelling real systems dynamics are generally done by representing it through coupled nonlinear systems of differential equations, but getting analytical solutions of such systems is most of the times not possible \cite{Strogatz}. Perturbation theory, linearization and numerical analysis techniques are the major tools to solve such coupled nonlinear differential equations. However, such techniques provide an approximate solutions of the system which sometimes provide us misleading results. $\dot{x}=-x $ has solution $x= exp(-t)$ and $x=0$ shows the asymptotic stable solution. if we change the coordinate to $z= logx$ and equation will be $\dot{z}=-1$ with solution $ z= -t$ and it's nowhere stable, behaviour of solution in two different coordinates are totally different. Such nonlinear system can't be fully describe by single coordinate system, rather it required numbers of compatible coordinate system(called charts) to describe the system. Therefore, it is crucial to define the system on a manifold that can provide more precise and clear outcomes compared to a system described in a local coordinates.

{\noindent}Differential geometry and Lie algebra provide a natural setting for solving nonlinear kind of differential equations. Where, Differential geometry provides a framework for analyzing the geometric properties of manifolds and their associated tangent spaces. Nonlinear systems can be represented as trajectories on these manifolds, and differential geometry techniques, such as the theory of differential forms, can be employed to study their dynamics and properties. Further, offers a powerful tool for analyzing the symmetries and transformations of nonlinear systems. it convert the problem of abstract algebra and differential equation into linear algebra which overwhelmingly reduce problem into simple form and can be easily solved \cite{Olver2000}.
\\

{\noindent}To steer the ecological system starting from an initial disease state $X_0$ to reach a desirable healthy state $X_D$, a certain set of control signals $u_j;j=1,2,...,m$ are applied to a set of suitable driver nodes in the ecological system network. The full set of control equations for the GLV model is then represented by following: \eqref{control}, 
\begin{eqnarray}
\label{control_eqn}
\frac{d\textbf{X}}{dt}&=&\textbf{F(X)}+\textbf{G(X)U}(t)\nonumber\\
&=&D[\textbf{X}]\left[\textbf{A}\textbf{X}+\textbf{r}\right]+\sum_{j=1}^{m}g_j(x_j)u_j(t)
\end{eqnarray}
where $\textbf{X} \in M$, and \textbf{F} and $g_{1}, ... ,  g_{m}$ are vector fields on smooth manifold $M$ and the inputs are $\textbf{U} \in \mathbb{R} ^{m}$. \\

\begin{itemize}
\item\textbf{Controllability and reachability:}
Thus the general controllability problem can be defined as steering the systems to reach the desired \textit{healthy state} $X_D$ from an initial \textit{disease state} $X_{0} \in M $ in finite time with appropriate inputs $u_{j}$. To steer the system to the desired state we need to check desired states' accessibility and hence prove its controllability \cite{Brockett2014, Brockett1973,Conte1999, Nijmeijer1990}. For checking the accessibility, we have to calculate the matrix \textbf{C} given by,
\begin{eqnarray}
\label{matrix}
\textbf{C}=[g_1, g_2, [g_1,F],[g_2,F],[g_1,[g_1,F]], [g_2,[g_2,F]]... ]
\end{eqnarray}
where, $[...]$ is the commutator, such that,
\begin{eqnarray}
[G, F](x)=\frac{\partial F}{\partial x} G(x)-\frac{\partial G}{\partial x} F(x)
\end{eqnarray}
are the Lie brackets of $F$ and $G$ vector fields. For checking controllability and accessibility we do the following analysis:

The general controllability problem stated as the existence of the sets of points starting from point $x_{0} \in M $ in finite time with appropriate inputs $u_{j}, j \in R^{m}$ to reach desired final \textit{healthy state} \cite{Angulo2019,Conte1999}. The ecological network dynamics with control input signals $u_j;j\in m$ is said to be controllable to reach from $X_0$ to $X_D$ if and only if the controllable matrix given by \eqref{matrix} has full rank, which is called Kalman's controllability rank condition \cite{KalmanR}, given by,
\begin{eqnarray}
\label{Kalman}
rank[C]=n
\end{eqnarray}
\item\textbf{Autonomous elements and accessibility:}
The control equation is given as:
\begin{equation}
\frac{d\textbf{X}}{dt}=\textbf{F(X)}+ \textbf{G(X)U}(t), \; \mbox{where}\; \; \textbf{X}(t) \in \mathbb{R}^{n} \; , \; \textbf{G} \in \mathbb{R}^{n \times m }, \; \; \; \textbf{U}(t) \in \mathbb{R}^{m} 
\label{auto}
\end{equation}
For eq. \ref{auto} we define autonomous elements a non-zero function $ \zeta(\textbf{X}(t)) $ such that there exist an integer \( p \leq n+1 \) such that $ \Phi(\zeta, \Dot{\zeta}, \Ddot{\zeta}, .... , \zeta^{p} )= 0 $ where $ \Phi $ is a meromorphic function( single valued analytic functions which allow Taylor series expansions). System is said to be accessible if it does not have any autonomous elements. Autonomous element are internal variables of the system and are not affected by system control\cite{Angulo2019,Conte1999} Therefore, control signals do not affect the evolution of $\zeta(\textbf{X}(t))$ and hence, the ecosystem \eqref{control_eqn} is defined to be accessible if $\zeta(\textbf{X}(t))$s are absent in it (details of proof can be seen in \cite{Angulo2019,Conte1999}).\\  

{\noindent}There is a formal way to identify autonomous elements in a given system\cite{Conte1999}. We define a differential field whose elements are spanned by vectors of meromorphic functions. We then construct Lie algebras and sub-algebras from these vector fields by iterative filtering of differential one-forms. The successive Lie algebras so created then give us a criteria for detecting autonomous elements in the given dynamical system. For our case we obtain the subspace $H_3 ={0}$ hence proving that there are no autonomous elements in the system \cite{Conte1999}.
\end{itemize}

\subsection{Numerical scheme with LQR}

{\noindent}When there are high-degree non-linearities as in our Model 2 , we cannot have closed form analytical solutions. We must make use of available computational tools for solving the corresponding optimization problem. We follow \cite{Angulo2019} in designing our linear model predictive control (MPC) using linear quadratic regulator (LQR). This algorithm aims to solve the impulsive control problem. The problem of finding control inputs sequence which are time-optimal is a variational problem requiring an optimization of a cost function under the constraints of given control-inputs.\\

{\noindent}The details of the numerical algorithm is given in the Appendix C as pseudocode. For this code, we first initialize all of the following quantities: $X_0$ is Initial state, to reach  $X_D$, the desired state, $\tau $  is the time between successive control interventions, n is number of species, m is the number of driver species or Input nodes, A is {$\mathbb{R}^{n \times n} $ } adjacency matrix of the species network, B is {$\mathbb{R}^{n \times m} $ } input matrix, Q  is {$\mathbb{R}^{n \times n}  $ } penalty matrix for deviations from $X_D$, R is the {$\mathbb{R}^{m \times m} $ } control effort matrix and $r$ is -$AX_0$ growth rate vector. All these parameters act as inputs to the programme which gives us the values of the control input sequences as output.\\
 
{\noindent}The programme is written in julia  using the package \textbf{dlqr} in \textbf{ControlSystems} \cite{julia} module and adapting the code for our requirements from \cite{angulo}. We have varied the parameters $\tau$, $Q$ and $R$ to give a general overview of how we can get various schemes of controls depending on these parameters which can be interpreted as various combinations of economic efforts, labour costs etc. 
\section{Results}

{\noindent}We present the results of analysis of the impact of possible control strategies on invasive plant \textit{Lantana camara}. We used our two basic minimal models (see the section \textit{Models and methods}) for this analysis to establish few possible strategies.

\subsection{The proposed models of \textit{Lantana camara} can be controlled with minimal number of control signals}
{\noindent}The first proposed model system in which \textit{L.camara} grows with self-sustained activity indicated by self-looped process with positive growth rate \textbf{r} and $A_{11}<0$ (negative value triggering to decease it's own population $x_1$) (see the \textit{Models and Methods} section). From the GLV equations \eqref{control_eqn} of the ecological network, we have the following function for the proposed model 1,
\begin{eqnarray}
\label{glv_m1}
\frac{d\textbf{X}}{dt}&=&\textbf{F(X)}+ \textbf{G(X)U}(t) \nonumber \\
&=&D[\textbf{X}]\left[\textbf{AX+r}\right]+\sum_{j=1}^{m}g_j(x_j)u_i\nonumber\\
\textbf{F}&=&\left[\begin{matrix}
-A_{11} x_{1}^{2}+r_{1} x_{1} \\
-A_{21} x_{2} x_{1}+r_{2} x_{2} \\
-A_{31} x_{3} x_{1}+A_{32} x_{3} x_{2}+r_{3} x_{3}
\end{matrix}\right]\\
&&\textbf{g}_1=\left[\begin{matrix}1\\0\\0\end{matrix}\right],~~
\textbf{g}_2=\left[\begin{matrix}0\\1\\0\end{matrix}\right],~~
\textbf{U}=\left[\begin{matrix}u_1\\u_2\end{matrix}\right]\nonumber
\end{eqnarray}
Now taking the parameter values taken for Model 1 as discussed in \textit{Models and Methods, Model 1} and applying two control signals $u_1$ and $u_2$ to the identified driver nodes $x_1$ and $x_2$, we constructed control matrix \textbf{C} using the procedure \eqref{matrix}. For checking the accessibility of a point i.e. if it can be reached through the two control inputs, we check the generic rank \cite{Brockett1973} of the accessibility distribution denoted by C 
\begin{eqnarray}
C &=&\left[\begin{matrix}
1 &0 &-1.2 x_{1}+r_{1}\\
0 &1& -0.5 x_{2}\\
0&0&-0.8 x_{3}
\end{matrix}\right]\nonumber\\
dim[C]&=&3\nonumber\\
rank[C]&=&3;~~\forall\left[\begin{matrix}x_1\\x_2\\x_3\end{matrix}\right]
\neq\left[\begin{matrix}0\\0\\0\end{matrix}\right]
\end{eqnarray}
Detail calculations are provided in Appendix A.(a). This proves that Model 1 is controllable using the two inputs $u_1$ and $u_2$ as given in Fig \ref{fig3}c.\\

{\noindent}We analyse how the signals $u_1$ and $u_2$ control the dynamics of the \textit{Lantana camara}. For this the model system is allowed to evolve with time starting from an initial state $X_0$ with given input signals (impulsive) incorporated at their respective chosen nodes. In order to reach the desired state, we integrate the full equations \ref{m1_1}, \ref{m1_2}  \ref{m1_3}, and the details are shown in Appendix A b. On solving these coupled ODEs we obtain:
\begin{eqnarray}
x_{1}(t) &=&\frac{0.9 e^{0.54 t}}{\alpha+e^{0.54 t}} \label{mm1_s1} \\
x_{2}(t)&\approx &\frac{\beta e^{0.45 t}}{\left(\alpha+e^{0.54 t}\right)} \label{mm1_s2}\\
x_{3}(t_1)&=&  \frac{x_{30}(1+\alpha)^{1.333}}{\left(\alpha+e^{0.54 t_1}\right)^{1.333}}  \exp \left[0.65 t_1 +1.55\frac{\beta}{\alpha}\left[\theta_{t_{1}} e^{0.45 t_{1}}-\theta_{\alpha}\right]  \right]   \label{mm1_s3}\\
&& \alpha =\frac{0.9-\left(u_{11}+x_{10}\right)}{u_{11}+x_{10}} \nonumber\\
&&\beta = (x_{20}+u_{21})(1+\alpha)^{0.833}\nonumber\\
&&\theta_{\alpha} ={ }_{2} F_{1}\left(a, b, c,-\frac{1}{\alpha}\right), \, \; \theta_{t_{1}} ={ }_{2} F_{1}\left(a, b, c,-\frac{e^{0.54 t_{1}}}{\alpha}\right) \nonumber \\ 
&&a =0.8333 ,\,\; b =1 , \,\; c =1.8333 \nonumber
\end{eqnarray}
where $x_{10}$ is the initial state of \textit{Lantana camara}, $x_{20}$ initial state of C-plant, $x_{30}$ is the initial state of Microbes respectively. We implemented the control strategy in such a way that the first control input intervention time is at $t_0=0$ and the second intervention is applied after $\tau$ time that is at $t_1=t_0+\tau$. In this strategy, different input control signals are applied at first and interventions in such a way that $u_{11}$ and $u_{21}$ are applied at first intervention and $u_{12}$ and $u_{22}$ are applied at second intervention respectively. We have chosen the convention that $u_{11}$ and $u_{21}$ represent the amount of \textit{Lantana camara} plants removed from the soil in first and second interventions respectively and  $u_{21}$ and $u_{22}$ are amounts of C-plants planted in the space where \textit{Lantana camara} were removed.\\

{\noindent}In the Model-2, we incorporated self-loops in each node of the model network which allows to increase the number of nonlinear terms in the system's dynamics. Though self loops are introduced the driver nodes remain the same. Hence the number of control input signals are taken to be two which are incorporated in the same respective nodes as in Model-1. Now the GLV equations of this model can be obtained by using equation \ref{control} and equation \ref{data} as given below,

\begin{eqnarray}
\label{glv_m2}
\frac{d\textbf{X}}{dt}&=&\textbf{F(X)}+ \textbf{G(X)U}(t) \nonumber \\
&=&D[\textbf{X}]\left[\textbf{AX+r}\right]+\sum_{j=1}^{m}g_j(x_j)u_i\nonumber\\
\textbf{F}&=&\left[\begin{matrix}
A_{11} x_{1}^{2}+r_{1} x_{1} \\
A_{21} x_{2} x_{1}+A_{22}x^2_2+r_{2} x_{2}\\
A_{31} x_{3} x_{1}+A_{32} x_{3} x_{2}+A_{33}x^2_3+r_{3} x_{3}
\end{matrix}\right]\\
&&\textbf{g}_1=\left[\begin{matrix}1\\0\\0\end{matrix}\right],~~
\textbf{g}_2=\left[\begin{matrix}0\\1\\0\end{matrix}\right],~~
\textbf{U}=\left[\begin{matrix}u_1\\u_2\end{matrix}\right]\nonumber
\end{eqnarray}

We considered the parameter values taken as discussed in \textit{Models and methods, Model-2}. Next, we constructed control matrix \textbf{C} from the equation \eqref{glv_m2} using the procedure \eqref{matrix}. Then we checked the accessibility of the control system for the two control inputs by checking the generic rank of the \cite{Brockett1973} accessibility distribution matrix $C$ . We found the rank of matrix $C$ to be full rank given by,
\begin{eqnarray}
C&=&[g_1, g_2, [g_1,F],[g_2,F],[g_1,[g_1,F]], [g_2,[g_2,F]]... ]\\
C &=&\left[\begin{matrix}
1 &0 &-1.2 x_{1}+r_{1}\\
0 &1& -0.5 x_{2}\\
0&0&-0.8 x_{3}
\end{matrix}\right]\nonumber\\
dim[C]&=&3\nonumber\\
rank[C]&=&3;~~\forall\left[\begin{matrix}x_1\\x_2\\x_3\end{matrix}\right]
\neq\left[\begin{matrix}0\\0\\0\end{matrix}\right]
\end{eqnarray}
The detail calculations are given in Appendix B. This proves that Model-2 is controllable using the two inputs $u_1$ and $u_2$ as given in Fig \ref{fig3}c. Due to presence of second order non-linearities in equation \ref{m2_2} it is not possible to integrate it and obtain closed form solutions, hence we have only provided numerical solutions for Model-2.

\subsection{Control strategy of \textit{Lantana camara}}

{\noindent}The methods for managing growth of \textit{Lantana camara} using the control signals ($u_1$ and $u_2$) incorporated at the identified driver nodes in our model system is studied by analyzing the solutions of GLV equations given in equations \eqref{mm1_s1}, \eqref{mm1_s2} and \eqref{mm1_s3} for the Model-1. The necessary condition for the convergence of the Hyper-geometric function, ${ }_{2} F_{1}\left(a, b, c,-\frac{1}{\alpha}\right)$ in the equations \eqref{mm1_s1}-\eqref{mm1_s2} is given by the following:
\begin{eqnarray}
 u_{11} < 0.45-x_{10}
\end{eqnarray}
Further, from the convergence criteria of the function $ { }_{2} F_{1}\left(a, b, c,-\frac{e^{0.54 t_{1}}}{\alpha}\right) $, we also obtain a relation between the intervention time($\tau$) and the initial amount of Lantana present given as, $\tau <  \frac{ln(\alpha)}{0.54}$. This provides a quantitative estimate of the amount of input to be provided in the first intervention which should be dependent on the initial amount of Lantana Camara. 

We also obtain an estimate of the time-frame within which this first input must be applied. Further on substituting the value of $x_{3}\left(t_{1}\right)=x_{3 d}$ in \eqref{mm1_s3} we obtain the relation between the first intervention inputs as :
\begin{eqnarray}
u_{21}=\left[\frac{\ln \left(\frac{x_{3 d}\left[\alpha+e^{0.54 t_{1}}\right]^{1.33}}{x_{31}(\alpha+1)^{1.33}}\right)-0.65 t_{1}}{\frac{1.55(1+\alpha)^{0.833}}{\alpha}\left[\theta_{t_{1}} e^{0.45 t_{1}}-\theta_{\alpha}\right]}\right]-x_{20} 
\label{m1in}
\end{eqnarray}
This shows that the inputs are not independent quantities i.e. the ratio of Lantana removed and the amount of C-plant planted have a fixed ratio as given by \ref{m1in}. For second intervention we need to put 
$u_{12}= x_{1d}-x_1(t_1) $ and $u_{22}= x_{2d}-x_2(t_1)$. We have taken $X_D = [x_{1d},x_{2d},x_{3d}]^T=[0.1,0.8,0.6]^T $.\\

{\noindent} The initial state $X_0=[x_{10}~x_{20}~x_{03}]^T$, which was in \textit{disease state}, where, concentrations of the ecological species are unbalanced, be evolve with time to reach a \textit{healthy state} $X_D=[x_{1d}~x_{2d}~x_{3d}]^T$ as given in the equations \eqref{mm1_s1}, \eqref{mm1_s2} and \eqref{mm1_s3}. In \textit{healthy state}, all the constituting species (\textit{Lantana camara} and C-plant) should have equal participation in terms of concentrations of the respective species, which is termed as \textit{population balance condition}, to keep proper diversity in the ecosystem. Taking the ideal case of such state, such that, $x_1(t)=x_{2}(t)$ at time $t$, and substituting the equations \eqref{mm1_s1} and \eqref{mm1_s2} and simplifying the terms, we have,
\begin{eqnarray}
\label{pb}
\frac{u_{21}}{x_{20}}=\lambda\left[\frac{u_{11}}{x_{10}}+1\right]^{0.833}e^{0.09t}-1=\Lambda\left[\frac{u_{11}}{x_{10}},t\right]
\end{eqnarray}
where, $\displaystyle\lambda=0.9^{0.167}\left[\frac{x_{10}^{0.833}}{x_{20}}\right]$ is a constant. From the equation \eqref{pb}, whether the second control signal will be activator ($u_{21}>0$) or inhibitor ($u_{21}<0$) to the species $x_2$ depends on the ratio $u_{11}/x_{10}$ and time of application $t$. Now, we have the following strategy of applying the control signals at their respective driver nodes.\\

\noindent\textbf{Proposition 1} \textit{The transition from initial disease state to a healthy state can take place if the control signal applied on\textit{ Lantana camara }is inhibitor.}\\

\noindent\textbf{Proof:} \textit{The applied two control signals are in fact interrelated by the equation \eqref{pb}. Suppose, the nature of control signal $u_{21}$ is activator to enhance the growth of $x_2$ i.e. C-plant, then from the equation \eqref{pb}, the right hand side to be larger than zero for positive $u_{21}$. Now, simplifying the inequality, we have,}
\begin{eqnarray}
\label{act}
\frac{u_{11}}{x_{10}} > \frac{e^{-0.09t}}{\lambda^{1.2}}-1
\end{eqnarray}
\textit{From this equation \eqref{act}, we can easily see that $x_{10}>>x_{20}$ as we started from initial disease state, and hence $\lambda>>1$. Further, as $t$ increases $e^{-0.09t}$ decreases and is always less than one. Hence, the inequality \eqref{act} does not hold. This indicates that in order to keep $u_{21}$ as an activator, $u_{11}$ has to be inhibitor in order to have the population of \textit{Lantana camara} and C-plant balanced for keeping diversity. Further, for $u_{21}$ to be inhibitor, from the equation \eqref{pb}, we have,}
\begin{eqnarray}
\label{inh}
\frac{u_{11}}{x_{10}} < \frac{e^{-0.09t}}{\lambda^{1.2}}-1
\end{eqnarray}
\textit{which is always true when $x_{10}>>x_{20}$. Hence, in the case of the time evolution of ecosystem having invasive plant \textit{Lantana camara} and C-plant, control input to be applied to the \textit{Lantana camara}should always be inhibitor no matter $u_{21}$ is inhibitor or activator.}\\

\noindent\textbf{Proposition 2} \textit{Successive interventions are needed for proper control of\textit{ Lantana camara }in the ecosystem}\\

\begin{figure*}[htbp]
    \centering
    \includegraphics[scale=0.46]{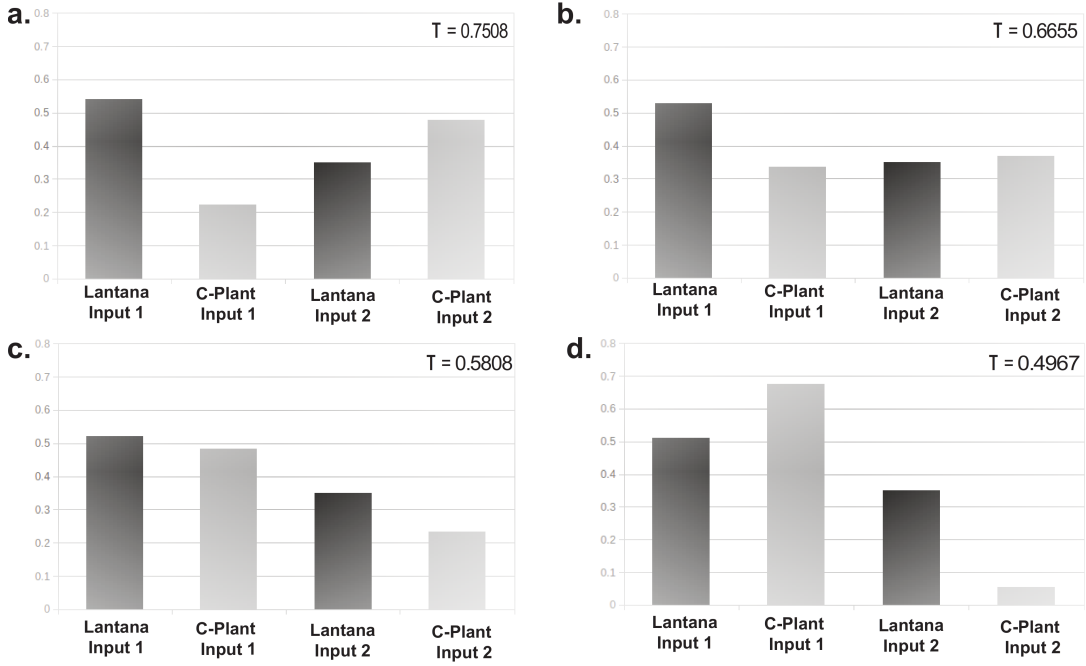}
    \caption{ Inputs for various schemes of interventions $\tau$ for Model 1. Input 1 refers to first intervention $u_{11},\;u_{21}$ and Input 2  refers to second interventions $u_{12},\; u_{22}$ on \textit{Lantana camara} and C-plant respectively.  }
    \label{fm1}
\end{figure*}

\begin{figure*}[htbp]
    \centering
    \includegraphics[scale=0.35]{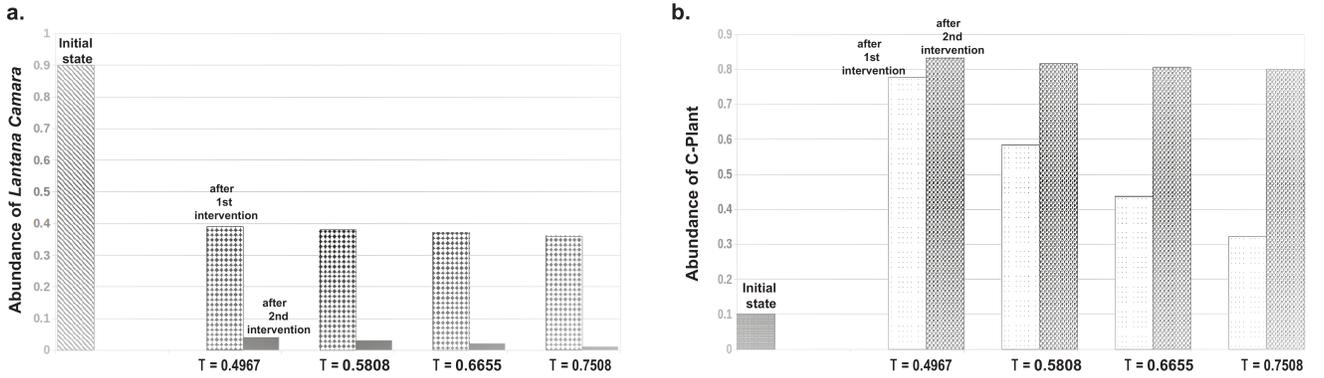}
    \caption{For Model 1, changes of abundances after giving control interventions for various $\tau$ schemes on (a) \textit{Lantana camara} (b) C-Plant. The initial states are shown with the bars marked "Initial state". The abundances after 1$^{st}$ intervention and 2$^{nd}$ intervention are marked above the bars. We have taken for all cases $X_D = [x_{1d},x_{2d},x_{3d}]^T=[0.1,0.8,0.6]^T $}. This shows that the proposed control can achieve the desired states within tolerable errors. 
    \label{fm2}
\end{figure*}

\noindent\textbf{Proof:} \textit{The equation \eqref{pb} indicates that bringing disease state to healthy depends on the ratio of initial condition in $\lambda$ ($x_{10}$ and $x_{20}$), ratio $\frac{u_{11}}{x_{10}}$, where, administration of input control signal at \textit{Lantana camara} can be optimized depending on the ratio of initial conditions to bring back to the healthy state quickly. The typical time of intervention could be $\displaystyle\tau <  \frac{ln(\alpha)}{0.54}$, such that,}

\begin{eqnarray}
 \alpha =\frac{0.9-\left(u_{11}+x_{10}\right)}{u_{11}+x_{10}}=\frac{\frac{0.9}{x_{10}}}{\frac{u_{11}}{x_{10}}+1}-1=\psi\left[\frac{u_{11}}{x_{10}}\right]
\end{eqnarray}
\textit{Hence, this ratio provides how much needs to wait for the next intervention and so on until we reach the healthy state of the ecosystem. This gives us the total control input sequence required to steer the system to our desired state. Hence, various control schemes can be devised according to the intervention time chosen. To prove it numerically, we simulate the model system (Model-1) with successive intervention times $\tau$. Some $\tau$ according to the bounds provided earlier have been shown in Figure \ref{fm1}(a-d). It shows the control-inputs for the two interventions applied to Lantana and C-plants. As we can see, the first intervention input to C-plant increases and the second intervention input to C-plant decreases with $\tau$ maintaining a trade-off between the time required and the amount of C-plant to be planted in replacement of Lantana. Figure \ref{fm2}(a-b) shows the output of the system to the above inputs schemes. All of the $\tau$ schemes perform within tolerable error limits. Figure \ref{fm2}(a) shows that Lantana has been reduced significantly after the second intervention. Thus we can see that we may not be able to eradicate Lantana but we can surely keep it in check and control.}\\
\begin{table*}[htbp]
    \centering
\begin{tabular}{|p{1.8cm}|| p{2cm} | p{2.0cm} | p{3.5cm} p{3.50cm} p {2.5cm}|}
    \hline
    \hfil \textbf{ Scheme($\tau$)} &  \hfil \textbf{ Effort ($\mathbf{Q}$)} &\hfil \textbf{ Penalty} ($\mathbf{r}$)  &\hfil \textbf{ Final abundance} & \hfil \textbf{ Impact(\%)} & \hfil \textbf{ Error }\\
    & & &\hfil \textbf{ of \textit{Lantana camara} } & \hfil \textbf{on \textit{Lantana camara}}&\\
    \hline
    \multirow{16}{*}{$\tau = 0.7508$} &\hfil
    \multirow{4}{*}{\hfil $Q_{1} = 0.1$} &\hfil $ 0.1$   &\hfil 0.4589 & \hfil 49.0075 & \hfil 0.3589 \\\cline{3-6}
    &\hfil&\hfil $ 0.01$   &\hfil  0.1895  &\hfil 78.9448 &\hfil 0.0895 \\ \cline{3-6}
    &\hfil&\hfil $ 0.001$   &\hfil 0.1454 &\hfil 83.8388 &\hfil 0.0454 \\ \cline{3-6}
    &\hfil&\hfil $ 0.0001$   &\hfil 0.1425 &\hfil 84.1801 &\hfil 0.0425 \\ \cline{2-6}\cline{2-6}\cline{2-6}&\hfil 
    \multirow{4}{*}{$Q_{2}=0.01$} &\hfil $0.1$   &\hfil 0.7414 &\hfil 17.6205 &\hfil 0.6414 \\\cline{3-6}
    &\hfil&\hfil $0.01$  \hfil &\hfil 0.4589 &\hfil 49.0075 &\hfil 0.3589 \\\cline{3-6}
    &\hfil&\hfil $0.001$   &\hfil 0.1894 &\hfil 78.9448 &\hfil 0.0894 \\\cline{3-6} 
    &\hfil&\hfil $0.0001$   &\hfil 0.1454 &\hfil 83.8388 &\hfil 0.0454 \\ \cline{2-6}\cline{2-6}\cline{2-6}
    &\hfil
    \multirow{4}{*}{$Q_{3}=0.001$} &\hfil $0.1$   &\hfil 0.8684 &\hfil 3.5050 &\hfil 0.7684 \\ \cline{3-6}
    &\hfil&\hfil $0.01$   &\hfil 0.7414 &\hfil 17.6202 &\hfil 0.6414 \\ \cline{3-6}
    &\hfil&\hfil $0.001$   &\hfil 0.4589 &\hfil 49.0075 &\hfil 0.3589 \\ \cline{3-6}
    &\hfil&\hfil $0.0001$   &\hfil 0.1894 &\hfil 78.9448 &\hfil 0.0894 \\ \cline{2-6}\cline{2-6}\cline{2-6}&\hfil
    \multirow{4}{*}{$Q_{4}=0.0001$} &\hfil $0.1$   &\hfil 0.8962 &\hfil 0.4124 &\hfil 0.7962\\\cline{3-6}
    &\hfil&\hfil $0.01$   &\hfil 0.8684 &\hfil 3.5050 &\hfil 0.7684 \\\cline{3-6}
    &\hfil&\hfil $0.001$   &\hfil 0.7414 &\hfil 17.6202 &\hfil 0.6414 \\\cline{3-6} 
    &\hfil&\hfil $0.0001$   &\hfil 0.4589 &\hfil 49.0075 &\hfil 0.3589 \\ 
    \hline    
    \end{tabular} 
    \caption{Performance of control schemes for different parameter values of effort($\mathbf{Q}$) and penalty($\mathbf{R}$) for $\tau = 0.7508$ which corresponds roughly to one month time intervals}
    \label{m2_t}
\end{table*}
\begin{figure*}[htbp]
    \centering
    \includegraphics[scale=0.38]{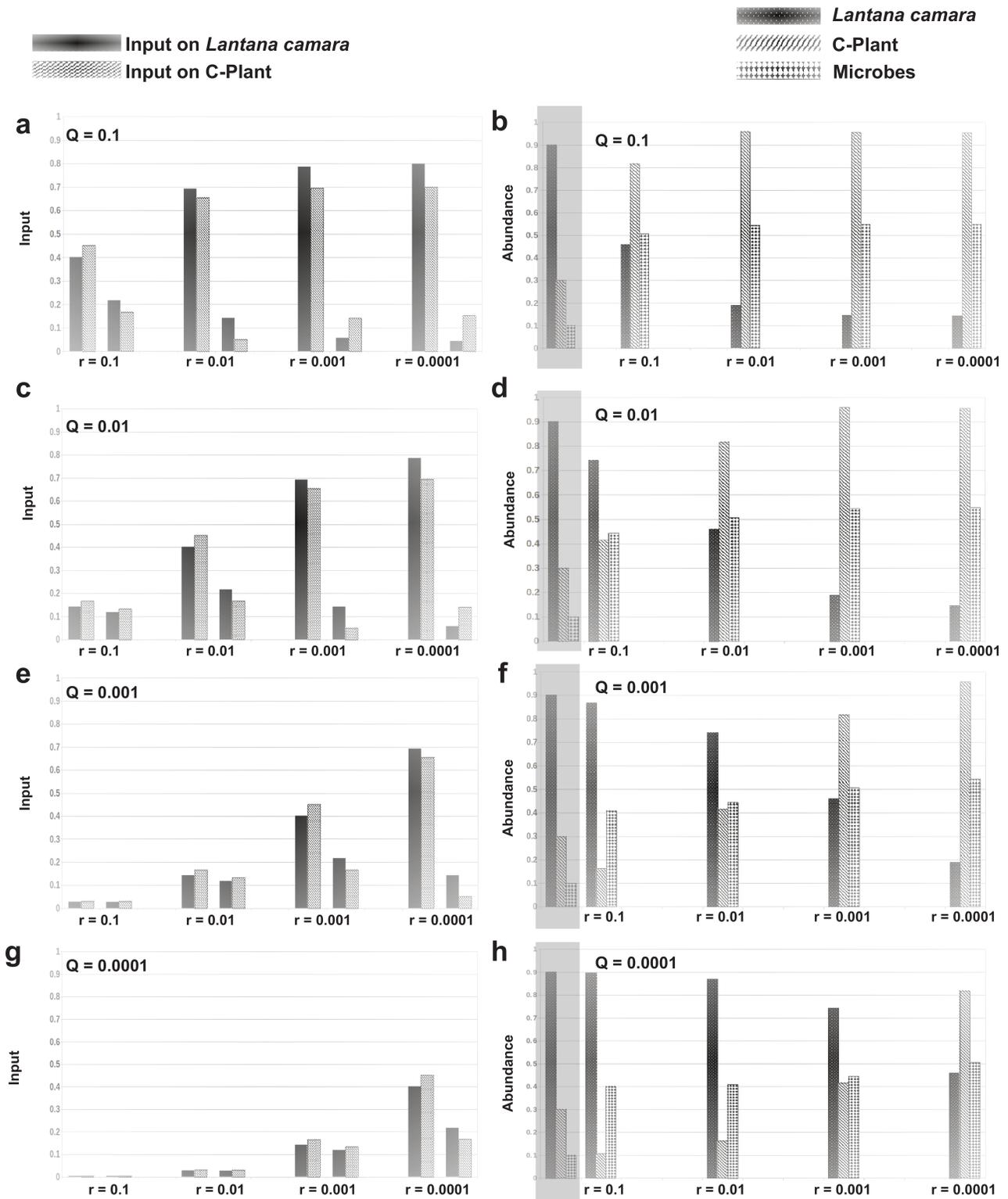}
    \caption{ Input and Output from schemes for tau = 0.7508 for various Q and R. (a,c,e,g) shows the input sequences i.e. first intervention $u_{11},\;u_{21}$ and second interventions $u_{12},\; u_{22}$ on $X_1$ and $X_2$ respectively. (b,d,f,h): Abundances after the interventions are applied. The bars in the shaded box are the initial abundances.  }
    \label{t7}
\end{figure*}
\subsection{The effect of nonlinearity in the ecosystem dynamics}

{\noindent}The nonlinearity in the system's dynamics plays various important roles in regulating the dynamics. We incorporate higher nonlinear terms by increasing the number of self-loops at the nodes of model network (Models and methods, Model-2). Because of the increase in the number of nonlinear terms in the model equations \eqref{glv_m2} is not analytically tractable. Hence, we applied LQR numerical method (Models and methods: numerical scheme with LQR) to simulate these coupled equations  \eqref{glv_m2} to study control.

{\noindent}We have solved control equations \eqref{glv_m2} numerically which incorporate optimization technique (see \textit{Models and methods}) for various intervention time $\tau$ schemes  with different effort($\mathbf{Q}$) and penalty ($\mathbf{R}$) parameters. We have chosen four intervention interval $\tau = 0.7508, 0.6655, 0.5808 \ \text{and} \ 0.4907 $ for each $\tau$ there corresponds four $\mathbf{Q}$ values and for each $\mathbf{Q}$ we have four $\mathbf{R}$ values. These $\tau$ values were chosen to coincide with the ones obtained analytically from Model 1 in order to obtain a comparative view of the efficiency of the MPC via LQR. These $\tau$ values also suggest the intervention time period to be chosen by the managing institution. \\

{\noindent}We have shown all input schemes for fixed $\tau = 0.7508$ which correpsond to a time scale of roughly 21 days between two interventions (according to our models) in panels \ref{t7}(a,c,e,g) and all corresponding output states of system in panels \ref{t7}(b,d,f,h) respectively. We have also highlighted the initial states of Lantana, c-plant and microbes in all the output panels for comparison purposes. We can see that not all control input sequences are efficient. We also get inferences about trade-off values for the amount of effort and the amount of penalty we impose for not reaching close to the desired state. For example consider comparing the schemes whose input sequence is shown in \ref{t7}(a) whose output is shown in \ref{t7}(b) and input sequence \ref{t7}(g) whose performance can be seen in \ref{t7}(h). Clearly the former \ref{t7}(a) with Q=0.1 is a better control strategy than the latter \ref{t7}(g) with Q=0.0001 for the given initial state of Lantana, c-plant and microbes as $[x_{10},x_{20},x_{30}]^T=[0.9,0.3,0.1]^T$ and $X_D = [x_{1d},x_{2d},x_{3d}]^T=[0.1,0.8,0.6]^T $. For the other $\tau$ schemes the impact of the control sequences are summarised in Fig. \ref{impact}.

{\noindent}We have provided the input-output performance for $\tau = 0.7508$ in the form of a Table \ref{m2_t} to get a general overview of the efficiency of the schemes. The figures for the rest of the input-output schemes Figure \ref{t6}, Figure \ref{t5} and Figure \ref{t4} have been provided in the appendix. Further, these nonlinear terms also depict a more realistic situation and thus makes control schemes as close to practical field operations as possible.  Hence, in devising sustainable management and making policy of controlling invasive plant \textit{Lantana camara}, we need to take into account all these parameters systematically.

\section{Conclusion and discussion}

{\noindent}We studied the character of the invasive plant \textit{Lantana camara} which affects various entities of the ecosystem e.g. species diversity. Even though this plant has some usefulness in terms of human usage, its negative impact is more prominent in the long run. In this work, we mainly focus is on the control of the outgrowth of this plant so that diversity of the ecosystem is properly maintained. In this perspective, we proposed two basic minimal models based on various characters of this plant and its interacting partners(represented as network) in the ecosystem through GLV dynamics of the constituting species abundance variables. For the control of the \textit{Lantana camara}, we incorporated control signals to driver nodes of the network and used control theory to investigate and propose strategies for managing it.\\

{\noindent}The study showed that there were minimal number of control signals to be incorporated to the driver nodes of proposed models, which satisfy Kalman's criteria of controllability, in order to control \textit{Lantana camara}. The dynamical equations are analyzed within the deterministic framework by solving the model equation analytically as well as numerically using optimization LQR algorithm. The solutions showed that control signals are dependent on initial species concentration also is quite important is to decide what type of control signals (inhibitor or activator) one needs to apply during the course of action. It is also found that during the control process is not a once-forever type action but sequential time-optimal interventions are essential and sufficient for managing \textit{Lantana camara}. Hence, we propose a technique involving successive interventions in order to get the healthy state within desirable time duration. \\

{\noindent}In dynamical systems, nonlinearity (degree of nonlinearity and/or number of nonlinear terms) plays very important roles to regulate and drive the system to various states. In our model systems, we found that increase in nonlinear terms in the system's dynamics, which are contributed from the self-loops in the network, facilitate the persistence of diversity in the ecosystem. Further, from the numerical solutions we observed that to control \textit{Lantana camara}, one needs to consider various factors, namely, the intervention time ($\tau$), values of various optimization parameters e.g. \textbf{R, Q} etc. to reach the desired state. Hence, we need to incorporate all these factors when we design control strategy policies.\\

{\noindent}On the other hand, there are various shortcomings of the theoretical models. The models we studied do not involve noise or stochastic aspects. This neglects various roles of noise in regulating the system. These probabilistic aspects can be factored in through various techniques and may give an edge to the optimal solutions so obtained. Further, efficient modelling, in theory, may give results which may not be feasible under the constraints of random changes in conditions attached to the control program e,g, climate changes, various coupling mechanisms etc. This happens when parameters which are kept constant in the model change randomly or large deviations occur. Thus, there exists  long debates about how the efficiency of the theoretical designs of controls in ecological problems do not reach the desired levels when applied to actual field operations. One solution to these standing questions is proposing integrated solutions based on Computer Aided Design (CAD) principles or GUI tools development (e.g. mobile application based solutions). These solutions can be in the paradigm of integrating computational tools, Human resourse management and operation research. One particular way of doing this would be creating digital twins (DT) of the system under considerations. Though DTs makes extensive use of Bayesian formalism, it can certainly help us be more adaptive and robust.\\
 
{\noindent}We reiterate that implementing such a control strategy is a hierarchical solution involving multiple agencies with multiple level of workforce skills and  efficiency. Identifying these numerous constraints at the designing stages of the plan would help in provided efficient end solutions. We know from game theory playing a slow or stagnant strategy of obtaining  marginal gains will not work in the case of an aggressively occupying Lantana. We need to be more non-complacent and more adaptive in this game i.e. changing strategies suited for the local contexts but not losing touch with the global goals. Our models and solutions are possibly one step into the direction of designing robust and manageable control strategies of \textit{Lantana camara}. The proposed models could be applicable to other invasive species dynamics as well. 
\vskip 1cm

\noindent\textbf{Author contribution}
SK, PM and RKBS conceptualised the model. SK and PM did the mathematical analysis and  simulations. RKBS supervised the work. All of the authors contributed in writing the manuscript.\\

\noindent\textbf{Acknowledgements} The authors would like to thank A Langlen Chanu, Jyoti Bhadana and Rubi Jain for valuable discussions.\\

\noindent\textbf{Funding} PM and SK would like to acknowledge JNU and UGC for financial support during the work. RKBS acknowledges DBT-COE for financial support.\\

\noindent\textbf{Conflict of interest:} The authors declare there is no conflict of interest.

\appendix

\section{Appendix For Model 1}
\subsubsection{Controllability (Accessibility)}
For controlling a dynamical system the generic rank of the C matrix must be $<=$ the dimension of the system.  
We calculated the controllability matrix(C) for the model 1 as given below: 
The dynamical system is given by :
\begin{gather}
\textbf{F}=\left[\begin{array}{l}
-A_{11} x_{1}^{2}+r_{1} x_{1} \\
-A_{21} x_{2} x_{1}+r_{2} x_{2} \\
-A_{31} x_{3} x_{1}+A_{32} x_{3} x_{2}+r_{3} x_{3}
\end{array}\right]
\end{gather}
The Jacobian is :

\begin{gather}
\frac{\partial \textbf{F}}{\partial x}=\left[\begin{array}{ccc}
-2 A_{11} x_{1}+r_{1} & 0 & 0 \\
-A_{21} x_{2} & -A_{21} x_{1}+r_{2} & 0 \\
-A_{31} x_{3} & A_{32} x_{3} & -A_{31} x_{1}+A_{32} x_{2}+r_{3} \\
\end{array}\right] 
\end{gather}

when we have substituted the values of $A_{ij}$ we get: \\

\begin{gather}
\frac{\partial \textbf{F}}{\partial x}=\left[\begin{array}{ccc}
-1.2 x_{1}+r_{1} & 0 & 0 \\
-0.5 x_{2} & -0.5x_{1}+r_{2} & 0 \\
-0.8 x_{3} & 0.7 x_{3} & -0.8x_{1}+0.7 x_{2}+r_{3} \\
\end{array}\right] 
\end{gather}

\begin{widetext}
\begin{gather}
{\left[g_{1}, \textbf{F}\right]=\left[\begin{array}{l}
-1.2 x_{1}+r_{1} \\
-0.5 x_{2} \\
-0.8 x_{3}
\end{array}\right]} 
{\left[g_{2}, \textbf{F}\right]=\left[\begin{array}{c}
0 \\
-0.5 x_{1}+r_{2} \\
0.7 x_{3}
\end{array}\right]} 
{\left[g_{1}\left[g_{1}, \textbf{F}\right]\right]=\left[\begin{array}{c}
-1.2 \\
0 \\
0
\end{array}\right]}
\end{gather}
\end{widetext}

The rest of the Lie brackets are need not be computed as all of them are zero. 
\begin{gather}
{\left[g_{2}\left[g_{2}, \textbf{F}\right]\right]=higher\,order \,Lie\,brackets=\left[\begin{array}{c}
0 \\
0 \\
0
\end{array}\right]} \\
C=[g_1, g_2, [g_1,\textbf{F}],[g_2,\textbf{F}],[g_1,[g_1,\textbf{F}]], [g_2,[g_2,\textbf{F}]]... ]\
\end{gather}

This yields the controllability matrix (C) as 
\begin{gather}
C =\left[\begin{array}{ccccc}
1 &0 &-1.2 x_{1}+r_{1}&0&-1.2\\
0 &1& -0.5 x_{2}&-0.5 x_{1}+r_{2}&0\\
0&0&-0.8 x_{3}&0.7 x_{3}&0
\end{array}\right] \\
C =\left[\begin{array}{ccc}
1 &0 &-1.2 x_{1}+r_{1}\\
0 &1& -0.5 x_{2}\\
0&0&-0.8 x_{3}
\end{array}\right] \\
\forall [x_1,x_2,x_3]^T \neq [0,0,0]^T \implies
\operatorname{dim}[C]=3
\end{gather}

where $C$ is the accessibility distribution

\subsubsection{Analytical solutions}
On solving the ODEs\ref{model1}we obtain :
\setlength{\jot}{3ex}
\begin{gather}
\int_{ u_{11}+x_{10}} ^{x_1(t)}  \frac{dx_1 }{-0.6x_1^2+0.54x_1} = \int_{0}^{t} dt \\
\implies x_{1}(t) =\frac{0.9 e^{0.54 t}}{\alpha+e^{0.54 t}} \label{m1_s1}  \\
\alpha =\frac{0.9-\left(u_{11}+x_{10}\right)}{u_{11}+x_{10}} \nonumber
\end{gather}
where $x_{10}$ is the initial state of \textit{Lantana camara},$u_{11}$ is the first intervention applied to \textit{Lantana camara}.

\noindent Similarly substituting \ref{m1_s1} in \ref{m1_2} and using separation of variables we have :

\begin{gather}
\int_{ u_{21}+x_{20}} ^{x_2(t)}  \frac{dx_2}{x_2} = \int_{0}^{t}  \left[\frac{-0.45e^{0.54t}}{\alpha +e^{0.54t} }+ 0.45 \right] dt\\
\implies x_{2}(t)=\frac{\beta e^{0.45 t}}{\left(\alpha+e^{0.54 t}\right)^{0.83}}  \\
\beta = (x_{20}+u_{21})(1+\alpha)^{0.833} \nonumber
\end{gather}

Here we make an approximation:
\begin{gather}
 x_{2}(t)\approx \frac{\beta e^{0.45 t}}{\left(\alpha+e^{0.54 t}\right)} \label{m1_s2}
\end{gather}
The error in doing so is within tolerance for the system considered here. \\
Substituting \ref{m1_s1}, \ref{m1_s2} in \ref{m1_3} and separating the variables we have: 

\begin{gather}
\setlength{\jot}{10pt}
\int_{ x_{30}} ^{x_3(t_1)}  \frac{dx_3}{x_3} = \int_{0}^{t_1} \left[ - \frac{0.72e^{0.54t}}{\alpha +e^{0.54t} }+\frac{0.7 \phi e^{0.45t}}{\alpha +e^{0.54t}}+ 0.65 \right] dt \nonumber \\
\begin{split}
\implies x_{3}(t_1)=  \frac{x_{30}(1+\alpha)^{1.333}}{\left(\alpha+e^{0.54 t_1}\right)^{1.333}}    \exp  \big[0.65 t_1\\+1.55\frac{\beta}{\alpha}\left(\theta_{t_{1}} e^{0.45 t_{1}}-\theta_{\alpha}\right) \big] \label{m1_a3} 
\end{split}
\\ \theta_{\alpha} ={ }_{2} F_{1}\left(a, b, c,-\frac{1}{\alpha}\right) \nonumber \\ 
\theta_{t_{1}} ={ }_{2} F_{1}\left(a, b, c,-\frac{e^{0.54 t_{1}}}{\alpha}\right) \nonumber \\ 
a =0.8333 ,\,\; b =1 , \,\; c =1.8333 \nonumber
\end{gather}

\section{Appendix For Model 2}
The dynamical system is :
\begin{gather}
\textbf{F}=\left[\begin{array}{l}
-A_{11} x_{1}^{2}+r_{1} x_{1} \\
-A_{21} x_{2} x_{1}-A_{22}x^2_2+r_{2} x_{2} \\
-A_{31} x_{3} x_{1}+A_{32} x_{3} x_{2}-A_{33}x^2_3+r_{3} x_{3}
\end{array}\right]
\end{gather}

The Jacobian is :
\begin{widetext}
\begin{gather}
\frac{\partial \textbf{F}}{\partial x}= \left[\begin{array}{ccc}
-2 A_{11} x_{1}+r_{1} & 0 & 0 \\
-A_{21} x_{2} & -A_{21} x_{1}-2A_{22}x_2+r_{2} & 0 \\
-A_{31} x_{3} & A_{32} x_{3} & -A_{31} x_{1}+A_{32} x_{2}-2A_{33}x_3+r_{3} \\
\end{array}\right] 
\end{gather}
\end{widetext}

On substituting values of $A_{ij}$ we get :
\begin{widetext}
\begin{gather}
\frac{\partial \textbf{F}}{\partial x}=\left[\begin{array}{ccc}
-1.2 x_{1}+r_{1} & 0 & 0 \\
-0.5 x_{2} & -0.5x_{1}-0.4x_2+r_{2} & 0 \\
-0.8 x_{3} & 0.7 x_{3} & -0.8x_{1}+0.7 x_{2}-16x_3+r_{3} \\
\end{array}\right] 
\end{gather}
\end{widetext}

We calculated the Lie Brackets as:
\begin{gather}
{\left[g_{1}, \textbf{F}\right]=\left[\begin{array}{l}
-1.2 x_{1}+r_{1} \\
-0.5 x_{2} \\
-0.8 x_{3}
\end{array}\right]} \\
{\left[g_{2}, \textbf{F}\right]=\left[\begin{array}{c}
0 \\
-0.5x_{1}-0.4x_2+r_{2} \\
0.7 x_{3}
\end{array}\right]} \\
{\left[g_{1}\left[g_{1}, \textbf{F}\right]\right]=\left[\begin{array}{c}
-1.2 \\
0 \\
0
\end{array}\right]} \\
{\left[g_{2}\left[g_{2}, \textbf{F}\right]\right]=\left[\begin{array}{c}
0 \\
-0.4 \\
0
\end{array}\right]} \\
{higher\,order \,Lie\,brackets=\left[\begin{array}{c}
0 \\
0 \\
0
\end{array}\right]}
\end{gather}

The accessibility distribution is given by :
\begin{gather}
C=[g_1, g_2, [g_1,\textbf{F}],[g_2,\textbf{F}],[g_1,[g_1,\textbf{F}]], [g_2,[g_2,\textbf{F}]]... ]\\
C =\left[\begin{array}{cccccc}
1 &0 &-1.2 x_{1}+r_{1}&0&-1.2&0\\
0 &1& -0.5 x_{2}&-0.5 x_{1}-0.4x_2+r_{2}&0&-0.4\\
0&0&-0.8 x_{3}&0.7 x_{3}&0&0
\end{array}\right] \\
C =\left[\begin{array}{ccc}
1 &0 &-1.2 x_{1}+r_{1}\\
0 &1& -0.5 x_{2}\\
0&0&-0.8 x_{3}
\end{array}\right] \\
\forall [x_1,x_2,x_3]^T \neq [0,0,0]^T \implies
\operatorname{dim}[C]=3
\end{gather}

\section{Algorithm used in LQR}

%

\begin{figure*}[htbp]
    \centering
    \includegraphics[scale=0.2635]{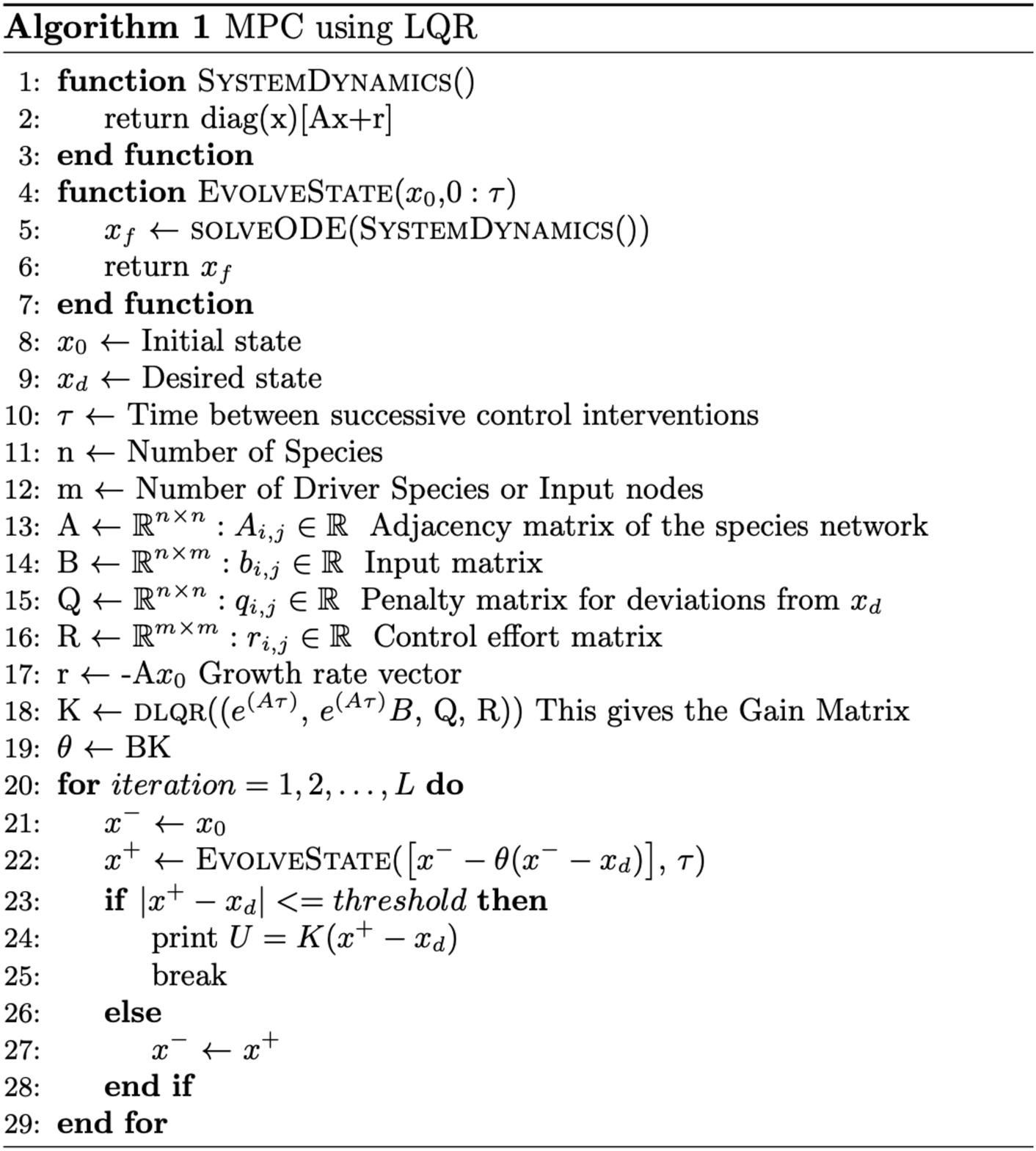}
    \caption{  }
    \label{i}
\end{figure*}

\begin{figure*}[htbp]
    \centering
    \includegraphics[scale=0.2635]{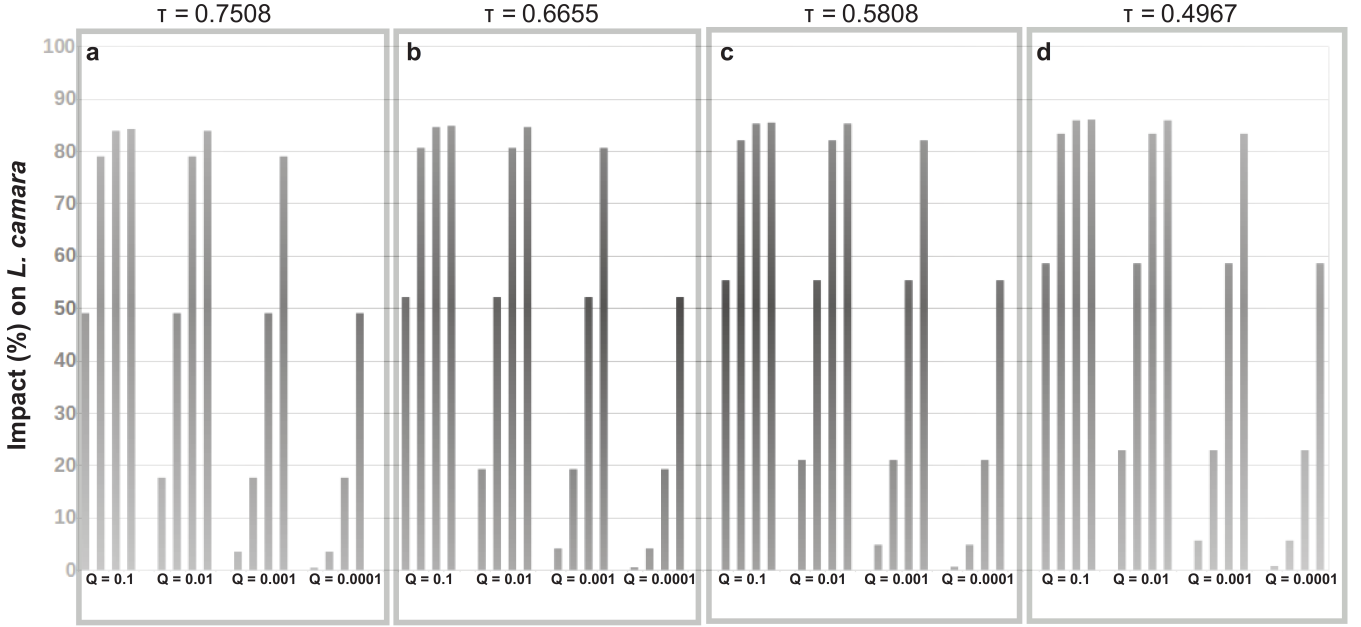}
    \caption{ The impact on abundance of \textit{Lantana camara} due to various control schemes }
    \label{impact}
\end{figure*}
\begin{figure*}[htbp]
    \centering
    \includegraphics[scale=0.38]{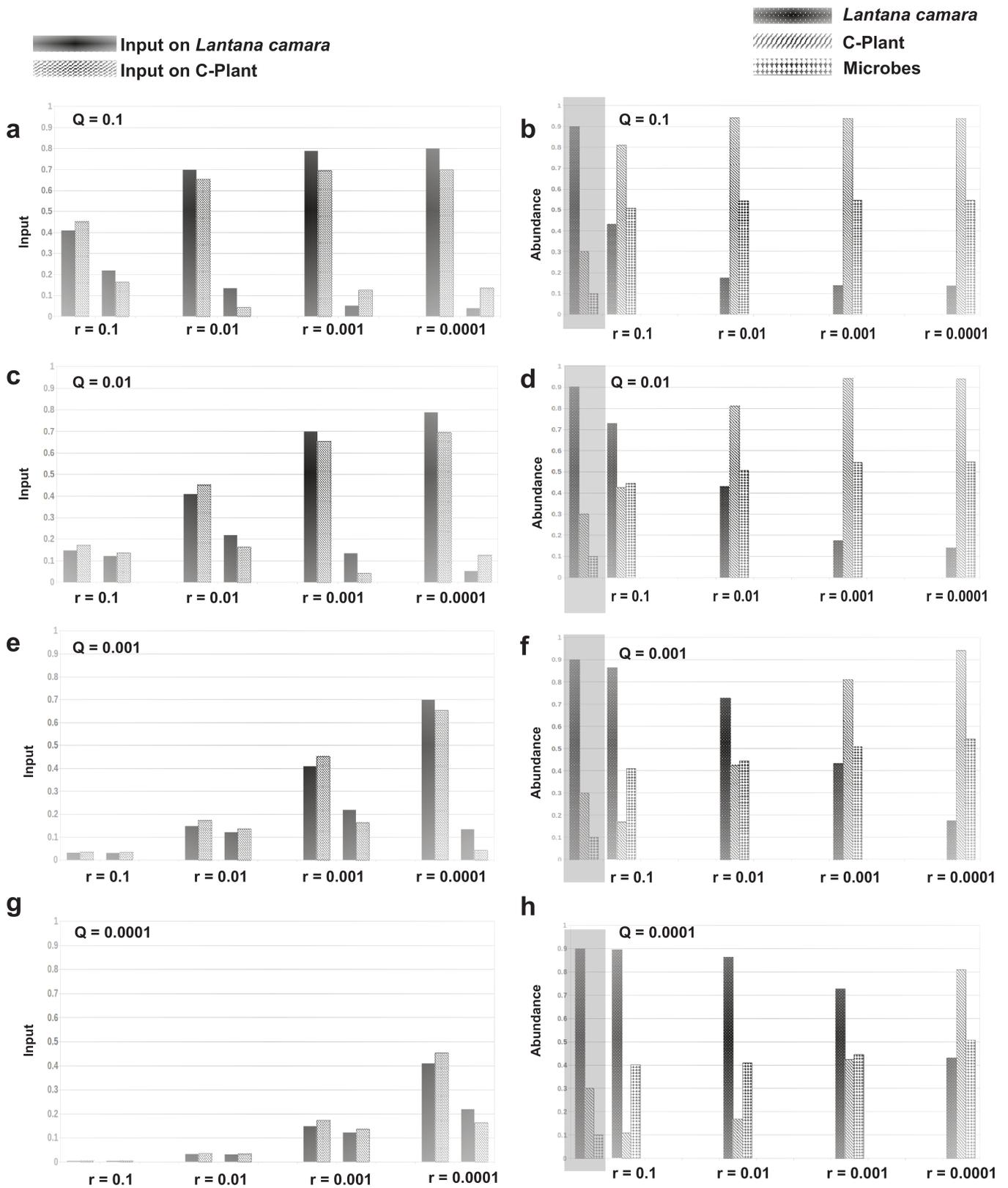}
    \caption{ Input and Output from schemes for $\tau$ = 0.6655}
    \label{t6}
\end{figure*}
\begin{figure*}[htbp]
    \centering
    \includegraphics[scale=0.38]{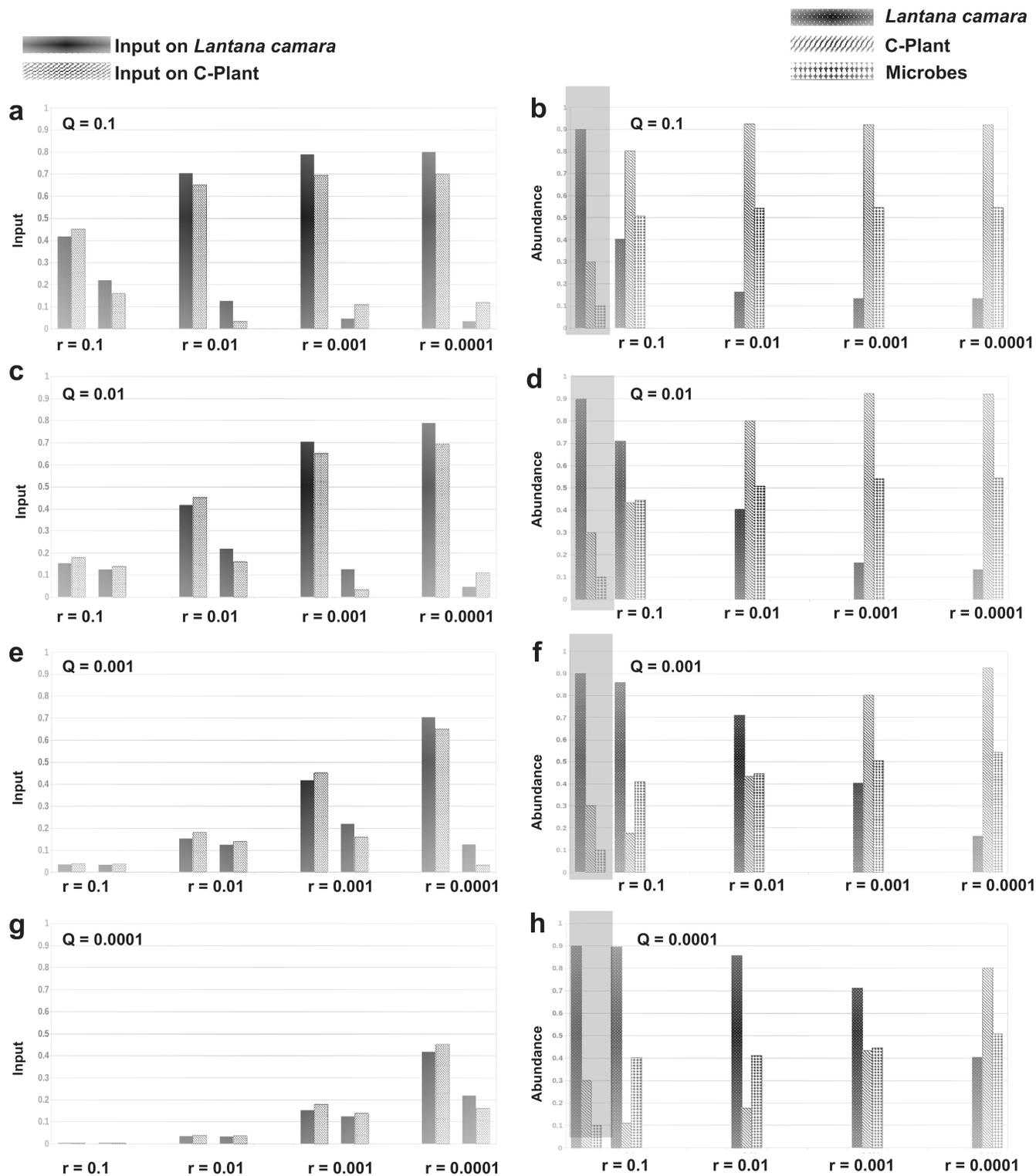}
    \caption{ Input and Output from schemes for $\tau$ = 0.5808}
    \label{t5}
\end{figure*}
\begin{figure*}[htbp]
    \centering
    \includegraphics[scale=0.38]{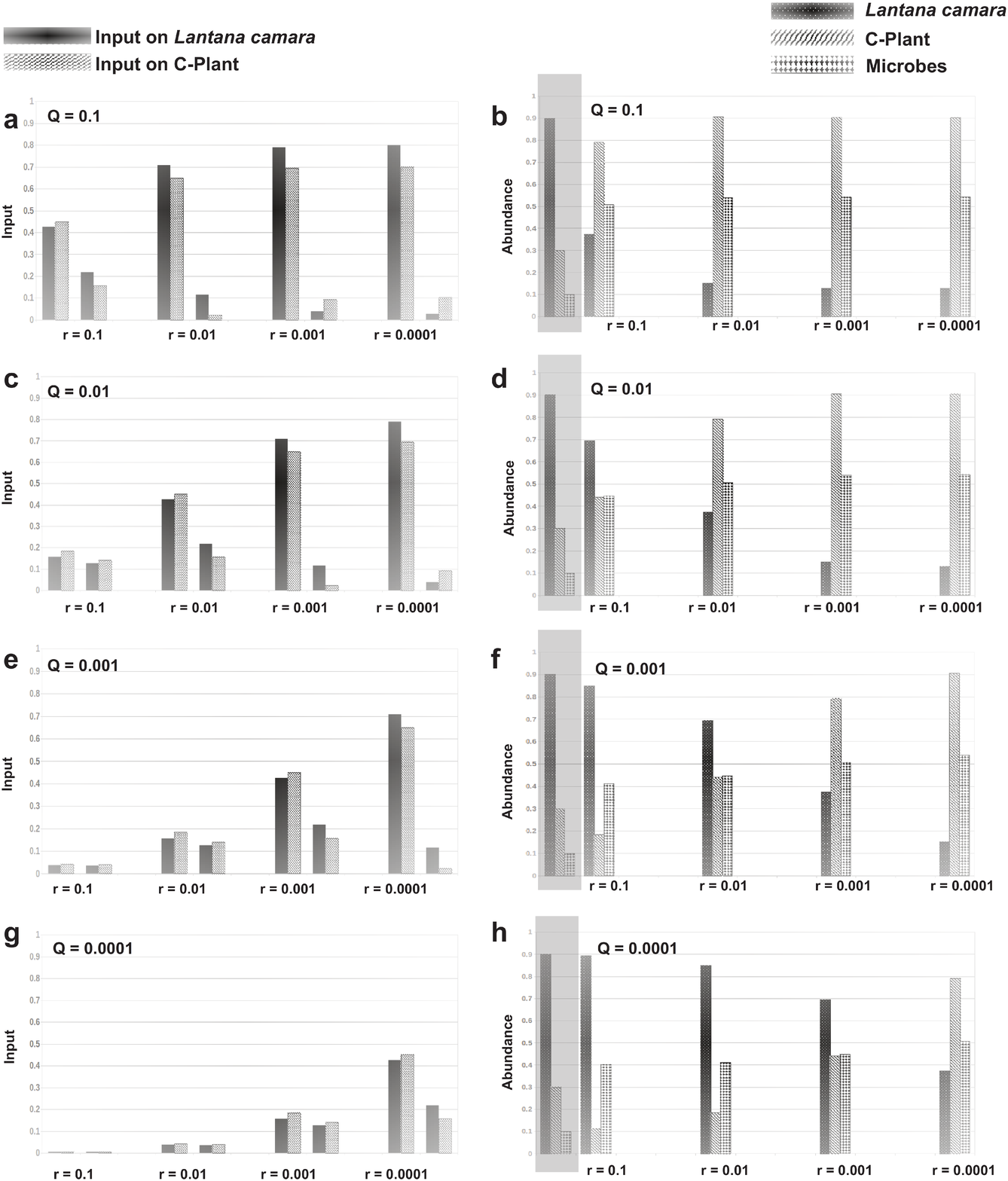}
    \caption{ Input and Output from schemes for $\tau$ = 0.4967}
    \label{t4}
\end{figure*}


\begin{thebibliography}{}

\bibitem{UNSDG}UN Sustainable Development Goals\url{https://sdgs.un.org/goals/goal15}
\bibitem{United}United Nations General Assembly Transforming our World: the 2030 Agenda for Sustainable Development A/RES/70/1 (United Nations, 2015).
\bibitem{Harris}Harris, J. M. (2003). Sustainability and sustainable development. International Society for Ecological Economic Internet Encyclopaedia of ecological economics. Retrieved from http://isecoeco.org/pdf/susdev.pdf. Date Accessed: 15 January 2019.
\bibitem{Alexander}Alexander JM, Atwater D, Colautti RI, and Hargreaves AL. Effects of species interactions on the evolution of species range limits. Phil.Trans.  R.  Soc.  B377: 20210020 (2022).
\bibitem{Malthus}Malthus, Thomas Robert. An Essay on the Principle of Population, London: Ward Lock 1890.
\bibitem{Sivakumar2018}Sivakumar, K., Rawat, G. S., Badola, R., Adhikari, B. S.,  Kamalakannan, B. (2018). Study on ecological socio-economic impact of invasive species, Prosopis juliflora and Lantana camara, and their removal from forest, common and fallow land of Tamil Nadu. Research Reports, Wildlife Institute of India \url{https://wii.gov.in/research_report2018}

\bibitem{Garkoti2021} Kumar, M., Kumar, S., Verma, A. K., Joshi, R. K.,  Garkoti, S. C. (2021). Invasion of Lantana camara and Ageratina adenophora alters the soil physico-chemical characteristics and microbial biomass of chir pine forests in the central Himalaya, India. Catena, 207, 105624.

\bibitem{Chauhan2022}Chauhan, S., Yadav, G., and Babu, S. (2022). Ecological Networks in Urban Forest Fragments Reveal Species Associations between Native and Invasive Plant Communities. Plants, 11(4), 541.

\bibitem{Kohli2006} Kohli, R. K., Batish, D. R., Singh, H. P.,  Dogra, K. S. (2006). Status, invasiveness and environmental threats of three tropical American invasive weeds (Parthenium hysterophorus L., Ageratum conyzoides L., Lantana camara L.) in India. Biological Invasions, 8(7), 1501–1510. \url{https://doi.org/10.1007/s10530-005-5842-1}

\bibitem{Negi2019} Negi, G.C.S., Sharma, S., Vishvakarma, S.C. et al (2019). Ecology and Use of Lantana camara in India. Bot. Rev. 85, 109–130 . \url{https://doi.org/10.1007/s12229-019-09209-8}

\bibitem{Vasudevan1991}Vasudevan, P., Jain S.K.(1991) Utilization of exotic weeds: an approach to control. In: Ramakrishnan P.S. (Ed.) Ecology of Biological Invasion in the Tropics. pp. 157-175. International Scientific Publishers, New Delhi, India 

\bibitem{Hiremath2018} Hiremath, A. J., Ayesha, P., Bharath, S. (2018). Restoring Lantana camara invaded tropical deciduous forest: the response of native plant regeneration to two common Lantana removal practices. Indian Forester, 144(6), 545-552.

\bibitem{Sharma2000}Sharma, R. P.,  Verma, T. S. (2000). Effect of long-term addition of lantana biomass on crop yields and N uptake in rice-wheat cropping in Himalayan acid Alfisols. Tropical agriculture, 77(2), 71-75.

\bibitem{Joshi1991}Joshi, S. (1991). Biological control of parthenium hysterophorus L. (asteraceae) by cassia uniflora mill (leguminosae), in Bangalore, India. Tropical Pest Management, 37(2), 182–184. \url{https://doi.org/10.1080/09670879109371572}

\bibitem{Sundaram2012} Sundaram B and Hiremath AJ (2012) Lantana camara invasion in a heterogeneous landscape: patterns of
spread and correlation with changes in native vegetation. Biol Invasions 14: 1127–1141

\bibitem{Ramaswami2014}Ramaswami, G., Prasad, S., Westcott, D., Subuddhi, S. P.,  Sukumar, R. (2014). Addressing the management of a long-established invasive shrub: the case of Lantana camara in Indian forests. Indian For, 140(2), 129-136.

\bibitem{Love2009}Love, A., Babu, S., Babu, C. R. (2009). Management of Lantana, an invasive alien weed, in forest ecosystems of India. Current Science, 1421-1429.

\bibitem{Flory2009}Flory, S. L.,  Clay, K. (2009). Invasive plant removal method determines native plant community responses. Journal of Applied Ecology, 46(2), 434-442.

\bibitem{Babu2009}Babu, S., Love, A.,  Babu, C. R. (2009). Ecological restoration of lantana-invaded landscapes in Corbett Tiger Reserve, India. Ecological Restoration, 27(4), 467-477.


\bibitem{Weidlich}Weidlich, E.W.A., Flórido, F.G., Sorrini, T.B., Brancalion, P.H.S.. Controlling invasive plant species in ecological restoration: a global review. J. Appl. Ecol. 57, 1806–1817 (2020).

\bibitem{Negi}Negi GCS, Sharma S, Vishvakarma SCR et al. Ecology and use of Lantana camara in India. Bot Rev 85:109–130 (2019).

\bibitem{Sharma}Sharma M, Alexander A, Saraf S et al Mosquito repellent and larvicidal perspectives of weeds Lantana camara L. and Ocimum gratissimum L. found in central India. Biocatal Agric Biotechnol 34, 102040 (2021). 

\bibitem{Arunkumar}Arunkumar B, Jeyakumar SJ, Jothibas M. A sol-gel approach to the synthesis of CuO nanoparticles using Lantana camara leaf extract and their photo catalytic activity. Optik (stuttg) 183:698–705 (2019).

\bibitem{Cohen1968}Cohen, J. E. (1968). Interval graphs and food webs: a finding and a problem. RAND Corporation Document, 17696.

\bibitem{Roberts1978}Roberts, F. S. (1978). Food webs, competition graphs, and the boxicity of ecological phase space. Theory and Applications of Graphs, 477–490. doi:10.1007/bfb0070404

\bibitem{Pickett}Pickett, S.T.A., Cadenasso, M.L., 2002. The ecosystem as a multidimensional concept:
meaning, model, and metaphor. Ecosystems 5, 1e10.
\bibitem{Mambuca}A. M. Mambuca, C. Cammarota, and I. Neri. Dynamical systems on large networks with predator-prey interactions are stable and exhibit oscillations. Phys. Rev. E, 105:014305, Jan 2022.

\bibitem{May1975}May, R. M.,  Leonard, W. J. (1975). Nonlinear Aspects of Competition Between Three Species. SIAM Journal on Applied Mathematics, 29(2), 243–253. \url{https://doi.org/10.1137/0129022}

\bibitem{Maier2013}Maier R. S. (2013) The integration of three-dimensional Lotka–Volterra systems. Proc. R. Soc. A 469 20120693
\url{http://doi.org/10.1098/rspa.2012.0693}

\bibitem{Ulanowicz}R. E. Ulanowicz. Quantitative methods for ecological network analysis. Comp. Biol. Chem. 28, 321–339 (2004).

\bibitem{Montoya}Montoya J. M., Pimm S. L., Sole R. V., Ecological networks and their fragility. Nature 442, 259 (2006).


\bibitem{Jones} Jones, H.P. et al. Restoration and repair of Earth’s damaged ecosystems. Proceed. Royal Soc. B 285, 20172577 (2018).
\bibitem{Singh}K. Singh, C. Byun and F. Bux. Ecological restoration of degraded ecosystems in India: Science and practices. Ecol. Eng. 182, 106708 ( 2022).




\bibitem{Angulo2019}Angulo, M. T., Moog, C. H., Liu, Y. Y. (2019). A theoretical framework for controlling complex microbial communities. Nature communications, 10(1), 1-12.

\bibitem{Liu2016}Liu, Y. Y.,  Barabási, A. L. (2016). Control principles of complex systems. Reviews of Modern Physics, 88(3), 035006.

\bibitem{Liu2011}Liu, Y. Y., Slotine, J. J., Barabási, A. L. (2011). Controllability of complex networks. nature, 473(7346), 167-173.

\bibitem{Lin1974} Lin, C. T. (1974). Structural Controllability. IEEE Transactions on Automatic Control, 19(3), 201–208. https://doi.org/10.1109/TAC.1974.1100557

\bibitem{Wei1963} Wei J.  and Norman E. (1963), "Lie Algebraic Solution of Linear Differential Equations", J. Math. Phys. 4, 575-581 \url{https://doi.org/10.1063/1.1703993 }



\bibitem{Conte1999} Conte, G., Moog, C. H.,  Perdon, A. M. (1999). Nonlinear control systems: An algebraic setting.

\bibitem{Brockett2014} Brockett, R. W. (2014). The early days of geometric nonlinear control. Automatica, 50(9), 2203-2224.

\bibitem{Brockett2015} Brockett, R. W. (2015). Finite dimensional linear systems. Society for Industrial and Applied Mathematics.

\bibitem{Brockett1973} Brockett, R. W. (1973). Lie algebras and Lie groups in control theory. In Geometric methods in system theory (pp. 43-82). Springer, Dordrecht.

\bibitem{Nijmeijer1990} Nijmeijer, H., Van der Schaft, A. (1990). Nonlinear dynamical control systems (Vol. 175). New York: Springer-verlag.



\bibitem{Olver2000}Olver, P. J. (2000). Applications of Lie groups to differential equations (Vol. 107). Springer Science Business Media.
 


\bibitem{Costanza1997}Costanza, V., Neuman, C. E. (1997). Managing cattle grazing under degraded forests: An optimal control approach. Ecological Economics, 21(2), 123–139. \url{https://doi.org/10.1016/S0921-8009(96)00098-5}


\bibitem{Costanza1990}Costanza, V., Neuman, C. E. (1990). Optimizing the process of forest fertilization as a control system. Fertilizer Research, 23(3), 151–164. \url{https://doi.org/10.1007/BF01073431}































\bibitem{Hejda}Hejda, M., Pysek, P. and Jarosik, V. Impact of invasive plants on the species richness, diversity and composition of invaded communities. J. Ecol., 97, 393-403 (2009).
\bibitem{Keyes}Keyes, A.A. et al. An ecological network approach to predict ecosystem service vulnerability to species losses. Nat. Commun. 12, 1586 (2021).
\bibitem{Pysek}Pysek P. and Richardson D.M. Invasive species, environmental change and management, and health. Annu. Rev. Envir. and Resor. 35, 25-55 (2010).
\bibitem{Sakai1954}Sakai, K. , Suzuki. Y (1954). Studies on competition in plants. III. Competition and spacing in one dimension. Jap. J. Genet. 29, 197-201


\bibitem{Doran}Doran JW, Zeiss MR. Soil health and sustainability: managing the biotic component of soil quality. Appl Soil Ecol. 15, 3-11 (2000).
\bibitem{Poole}Poole G.C. (2002) Fluvial landscape ecology: addressing uniqueness within the river continuum. Freshwater Biology, 47, 641–660.
\bibitem{Waldrop}Waldrop, M.P., Holloway, J.M., Smith, D.B., Goldhaber, M.B., Drenovsky, R.E., Scow, K.M., et al. The interacting roles of climate, soils, and plant production on soil microbial communities at a continental scale. Ecology 98: 1957-1967 (2017).
\bibitem{Pascale}Pascale, A., Proietti, S., Pantelides, I. S., and Stringlis, I. A. (2020). Modulation of the root microbiome by plant molecules: the basis for targeted disease suppression and plant growth promotion. Front. Plant Sci. 10:1741 (2020).
\bibitem{Bakker}Bakker MG, Schlatter DC, Otto-Hanson L, Kinkel LL. Diffuse symbioses: roles of plant–plant, plant–microbe and microbe–microbe interactions in structuring the soil microbiome. Mol Ecol. 23, 1571–83 (2014).
\bibitem{Hassani}Hassani, M. A., Duran, P., Hacquard, S. (2018). Microbial interactions within the plant holobiont. Microbiome 6, 58 (2018).
\bibitem{Kato}Kato-Noguchi, H.; Kurniadie, D. Allelopathy of Lantana camara as an Invasive Plant. Plants 10, 1028 (2021).
\bibitem{SinghHP}Singh, H.P.; Batish, D.R.; Dogra, K.S.; Kaur, S.; Kohli, R.K.; Negi, A. Negative effect of litter of invasive weed Lantana camara on structure and composition of vegetation in the lower Siwalik Hills, northern India. Environ. Monit. Assess. 186, 3379–33 (2014).


\bibitem{Hamilton}E. W. Hamilton, and D. W. Frank. Can plants stimulate soil microbes and their own nutrient supply? Evidence from a grazing tolerant grass. Ecology 82, 2397 (2001).
\bibitem{Wardle}D. A. Wardle et al., Ecological linkages between aboveground and below ground biota. Science 304, 1629–1633 (2004).

\bibitem{Fan2010}Fan, L., Chen, Y., Yuan, J. gang, Yang, Z. yi. (2010). The effect of Lantana camara Linn. invasion on soil chemical and microbiological properties and plant biomass accumulation in southern China. Geoderma, 154(3–4), 370–378. \url{https://doi.org/10.1016/j.geoderma.2009.11.010}



\bibitem{Newman}M. E. J. Newman. Networks: An Introduction. Oxford University Press (2010).

\bibitem{Jost}Christian Jost and Stephen P Ellner. Testing for predator dependence in predator-prey dynamics: a non-parametric approach. Proc. of the Royal Soc. of London B: Biol. Sc. 267(1453):1611–1620, (2000).

\bibitem{Goel}N. S. Goel, S. C. Maitra and E. W. Montroll. On the Volterra and Other Nonlinear Models of Interacting Populations. Rev. Mod. Phys. 43, 231–276 (1971).
\bibitem{Hassell}Hassell, M. P., Lawton, J. H. and May, R. M. Patterns of dynamical behaviour in single-species populations. J. Anim. Ecol. 45: 471-486 (1976).
\bibitem{Metz}Metz JAJ, Nisbet RM, Geritz SAH. How should we define “fitness” for general ecological scenarios. Trends Ecol. Evol. 7, 198–202 (1992).

\bibitem{Strogatz}Strogatz, S. (1994). Nonlinear Dynamics And Chaos: With Applications To Physics, Biology, Chemistry, And Engineering (Studies in Nonlinearity). Studies in nonlinearity. Perseus Books Group, 1 edition.
\bibitem{KalmanR}Kalman, R. E. Mathematical description of linear dynamical systems. J. Soc. Indus. Appl. Math. Ser. A 1, 152–192 (1963).
\bibitem{Miri}M. A. Miri, F. Ruesink, E. Verhagen, and A. Alù, Optical Nonreciprocity Based on Optomechanical Coupling, Phys. Rev. Appl. 7, 064014 (2017).

\bibitem{Hopcroft}Hopcroft, J. E. and Karp, R. M. An $n^{5/2}$ algorithm for maximum matchings in bipartite graphs. SIAM J. Comput. 2, 225–231 (1973).


\bibitem{Ravi2017}Ravichandran, K. R.,  Thangavelu, M. (2017). Role and influence of soil microbial communities on plant invasion. Ecological Questions, 27, 9–23. \url{https://doi.org/10.12775/EQ.2017.024}


\bibitem{Motte1978} Motte, L. G. (1978). The fifth international symposium on biological control of weeds. In Aquatic Botany (Vol. 5). https://doi.org/10.1016/0304-3770(78)90072-4



\bibitem{angulo} \url{https://github.com/mtangulo/DriverSpecies}

\bibitem{julia} Bagge Carlson, F., Fält, M., Heimerson, A., and Troeng, O. (2021). ControlSystems.jl: A Control Toolbox in Julia. In 2021 60th IEEE Conference on Decision and Control (CDC) IEEE Press. https://doi.org/10.1109/CDC45484.2021.9683403










\end{thebibliography}
\end{document}